\documentclass[12pt,preprint]{aastex}
\usepackage{epstopdf}

\def\exp#1{\times 10^{#1}}
\def\degr{^{\circ}}

\newcommand{\flux}{\rm erg\ s^{-1} cm^{-2}}

\newcommand{\kms}{{\rm km\ s^{-1}}}
\newcommand{\kmsmpc}{{\rm km\ s^{-1}Mpc^{-1}}}
\newcommand{\lsun}{\rm L_{\odot}}
\newcommand{\msun}{\rm M_{\odot}}
\newcommand{\mdyn}{M_{\rm dyn}}
\newcommand{\sfr}{\rm M_{\odot}\ yr^{-1}}
\def\h0{H_0}

\def\h100{h_{\rm 100}}

\newcommand{\lya}{Ly$\alpha$}
\newcommand{\ha}{H$\alpha$}
\newcommand{\hb}{H$\beta$}
\newcommand{\hi}{H~{\sc I}}
\newcommand{\hii}{H~{\sc II}}
\newcommand{\alii}{Al~{\sc II}}

\newcommand{\civ}{C~{\sc IV}}

\newcommand{\cii}{C~{\sc II}}
\newcommand{\ciii}{C~{\sc III}}
\newcommand{\feii}{Fe~{\sc II}}
\newcommand{\heii}{He~{\sc II}}
\newcommand{\mgii}{Mg~{\sc II} }
\newcommand{\oi}{O~{\sc I}}
\newcommand{\oii}{[O~{\sc II}]}
\newcommand{\oiii}{[O~{\sc III}]}
\newcommand{\siii}{Si~{\sc II}}
\newcommand{\siiii}{Si~{\sc III}}
\newcommand{\siiv}{Si~{\sc IV}}
\def\ni{N~{\sc I}}
\newcommand{\nv}{N~{\sc V}}

\slugcomment{Accepted for publication in the Astrophysical Journal}

\shorttitle{Dynamics of LBGs and Halos}

\shortauthors{Lowenthal et al.}

\begin{document}

\title{Dynamics of Lyman Break Galaxies and Their Host Halos}

\author{James D. Lowenthal}
\affil{Department of Astronomy, Smith College,  Northampton,
MA 01063; james@ast.smith.edu}

\author{David C. Koo}
\affil{UCO/Lick Observatory, University of California, Santa Cruz, CA
95064; koo@ucolick.org}

\author{Luc Simard}
\affil{Hertzberg Institute of Astrophysics, National Research Council
of Canada, Victoria, BC, V9E 2E7; Luc.Simard@nrc.ca}

\and

\author{Eelco van Kampen}
\affil{ESO, Karl-Schwarzschild-Str. 2, D-85748 Garching bei M\"unchen, Germany; evkampen@eso.org}

\begin{abstract}

We present deep two-dimensional spectra of 22 candidate and confirmed
Lyman break galaxies (LBGs) at redshifts $2<z<4$ in the Hubble Deep
Field (HDF) obtained at the Keck II telescope.  The targets were
preferentially selected with spatial extent and/or multiple knot
morphologies, and we used slitmasks and individual slits tilted to
optimize measurement of any spatially resolved kinematics.  Our sample
is more than one magnitude fainter and is at higher redshift than the
kinematic LBG targets previously studied by others.  The median target
magnitude was $I_{814}=25.3$, and total exposure times ranged from 10
to 50 ks.  We measure redshifts, some new, ranging from $z=0.2072$ to
$z=4.056$, including two interlopers at $z<1$, and resulting in a
sample of 14 LBGs with a median redshift $z=2.424$.  The morphologies
and kinematics of the close pairs and multiple knot sources in our
sample are generally inconsistent with galaxy formation scenarios
postulating that LBGs occur only at the bottom of the potential wells
of massive host halos; rather, they support ``collisional starburst''
models with significant major merger rates and a broad halo occupation
distribution.  For 13 LBGs with possible kinematic signatures, we
estimate a simple dynamical mass, subject to numerous caveats and
uncertainties, of the galaxies and/or their host dark matter halos.
Dynamical mass estimates of individual galaxies range from $4 \exp{9}
h^{-1} \msun $ to $ 1.1 \exp{11} h^{-1} \msun$ and mass estimates of
halos, based on close LBG pairs, range from $<10^{10} h^{-1}$ to $\sim
10^{14} h^{-1} \msun$ with a median value $1\exp{13}\msun$.
Comparison with a recent numerical galaxy formation model implies that
indeed the pairwise velocities might not reflect true dynamical
masses.  We compare our dynamical mass estimates directly to stellar
masses estimated for the same galaxies from SEDs, and find no evidence
for a strong correlation.  The diversity of morphologies and dynamics
implies that LBGs represent a broad range of galaxy or proto-galaxy
types in a variety of evolutionary or merger stages rather than a
uniform class with a narrow range of mass.

\end{abstract}


\section{Introduction}\label{sec:intro}

Lyman break galaxies (LBGs) are currently among our best windows on
the universe of galaxies at redshifts $z>2$.  LBGs are especially useful
for investigating galaxy formation and evolution because they are
relatively easy to find in large numbers (approaching 2 arcmin$^{-2}$
down to $R<25.5$; Steidel et al 2003; Giavalisco et al 2002) and
because they are relatively bright in the optical, thus permitting
optical spectroscopic followup.  Spectroscopic redshifts of several
thousand LBGs have been measured to date, with photometric redshifts
of tens of thousands also available \citep{pal07,ste04,ouc01,rav06}, and
luminosity functions beyond $z>4$ have been measured
\citep{kas06,bou06,iwa07,saw06}.  LBGs account for almost all of the
star formation at $z\sim3$ observable in optical windows, and roughly
half of the total star formation at those redshifts once dust-obscured
systems such as sub-mm galaxies are included \citep{sma02, cha05,
gia02}.  

Although a complete picture of LBGs and their relation to galaxies today remains elusive, many pieces of the picture are now available.  Spectra of LBGs clearly show evidence of strong bursts of star
formation, ranging from $1-100~~ \sfr$, confirmed with deep radio and
x-ray imaging with the VLA and {\it Chandra} by \citet{red04}, who
estimate a mean star formation rate (SFR) $\sim 50~~ \sfr$ for
UV-selected galaxies at $1.5<z<3$.  Deep images with HST
reveal small sizes $r_{1/2}\sim 4 h^{-1}$ kpc (with $h=H_0 / 72~ \kmsmpc$), high luminosities
$L\sim10^{12}\lsun$, and diverse morphologies, including multiple
small knots, diffuse halos, and asymmetrical linear features.  Deep {\it
Chandra} images detect only $\sim 3 \%$ of LBGs, implying that the AGN
fraction is at most that much \citep{lai05, lai06}.  

The spectra of LBGs also show evidence of strong outflows, e.g.
\lya\ emission lines red-shifted and interstellar
absorption lines blueshifted with respect to systemic redshifts \citep{sha03} and a paucity of QSO \hi\ absorption
lines near LBG sightlines \citep{ade03}, although \citet{des06} find
that only smaller outflow bubbles are needed to explain the spectra,
i.e. 0.5 Mpc comoving, implying that metals observed in the \lya\
forest are unlikely to come from LBGs.

Measuring and understanding the dynamical state and masses of LBGs is important for placing them in the context of galaxy formation.   A galaxy's mass affects its rate of
accretion of new material, its ability to retain gas against the
expulsive force of supernova winds, and possibly its eventual
morphological type.  The need to constrain LBG dynamics motivates the present work.

In the local universe, total galaxy mass is usually measured
kinematically by deriving rotation curves or emission line widths for
disk galaxies \citep{sof01}, measuring internal velocity dispersions
for elliptical and other spheroidal systems, and measuring velocity
distributions of satellite galaxies or other test particles for all
galaxy types \citep{zar93}, and then translating the observed velocity
field into a dynamical or virial mass.   The masses of larger systems such as galaxy groups and clusters, but not of their constituent galaxies, are likewise constrained by direct measurement of
their velocity dispersions, as well as by X-ray emission and gravitational lensing.

Three kinds of direct evidence to date have been used to study the dynamics of LBGs and LBG host halos: emission line widths and spatially-resolved kinematics (for total mass and dynamical state), spectral energy distributions (for stellar mass only), and clustering analysis. 

LBG mass measurements from redshifted \ha\ emission line widths and spatially-resolved kinematics generally fall in the range $0.5-25\exp{10} \msun$ \citep{pet01, erb04, erb06, for06, bou07, nes06, for09}.  Some  signs of rotation curves are seen at $z\sim2$ \citep{for06, for09, leh09}, and there is also evidence for superwinds, especially \lya\ emission lines blue-shifted with respect to stellar and interstellar absorption lines \citep{pet01, erb04, nes07}.  \citet{for09} found that one-third of 62 galaxies at $1.3 < z < 2.6$ they studied with the SINFONI integral field unit (IFU) on the Very Large Telescope (VLT) showed rotation-dominated kinematics, one-third are interacting or merging systems, and one-third are  dominated by  random motions; derived \ha\ dynamical masses for the whole sample are $3\exp{9} - 3\exp{11} \msun$, with median $M = 2.7\exp{10} \msun$.

Estimates of the stellar mass of LBGs have benefited from deep imaging with {\it Spitzer Space Telescope}'s Infrared
Array Camera (IRAC), including especially the GOODS and Extended Groth
Strip (EGS) fields.  Stellar masses $M_*$ of LBGs
at $z > 2$ generally range from $10^{9-11} \msun$ for luminosities $L>
{\rm L_*}$, with typical median values $M_* \sim 3\exp{10}\msun$
\citep{rig06, lab05, sha05, pap01, hua05}.  \lya\ emitting galaxies   (LAE's) show similar
stellar masses \citep{lai07}. 

Samples of LBGs in the redshift range $2<z<6$ are now large enough to
constrain star formation history in LBGs: the median LBG mass
is reported to be lower by around a factor of 10 at $z\sim6$ than at
$z\sim3$ \citep{ver07, lab06, yan06, eyl07}, implying significant
buildup of stellar mass through star formation and merging over that
time, and mirroring the observed increase by roughly a factor of 10
in the comoving stellar mass density $\rho_*$ from $z=3$ to $z=0$
measured in the distant red galaxy (DRG) population \citep{rud06, fon06}.
 
Finally, the observed clustering of LBGs has been used to constrain
the mass of dark matter halos in which the galaxies presumably reside.  Two-point correlation analysis reveals that LBGs cluster with each
other with typical correlation length $r_0 \sim 3-4 H_0^{-1}$ Mpc and implied halo masses $M_{\rm halo} \sim 10^{11-12} \msun$ \citep{gia01, kas06, ade05a, ouc01}, while still leaving
the masses of individual LBGs within those halos uncertain.  Clustering of LBGs
with damped \lya\ QSO absorption line clouds (DLAs) has also been
detected, with implied DLA halo masses on the order of
$2\exp{11}\msun$
\citep{bou03,bou04,bou05b,coo06}.

LBGs are also observed to cluster on small scales: \citet{col96} and
\citet{col97} found that the two-point correlation function of faint
objects in the HDF, consisting largely of LBGs at high redshift, peaks
between 0\farcs25 and 0\farcs4 ($\sim 1$ kpc at $z>1$) with
amplitude greater than 2, i.e. LBGs appear as multiple clumps of
emission rather than monolithic sources.  They interpret that
multiplicity as evidence of starbursting regions within otherwise
normal gas-rich galaxies.  Many of the targets discussed in the
current paper fall into that category of close pairs or multiple-knot
sources. 

Galaxy formation theorists have used observations of LBGs extensively
to check and revise their predictions and detailed descriptions of
mass assembly, star formation, feedback, morphology, and clustering
over cosmic time.  Hydrodynamic $N$-body simulations
\citep{cev09, fin06,wei02,mo99, nig06} and semi-analytic models (SAMs; Cole et al
2000; Somerville et al 2001; Bower et al 2006) of cold dark matter
(CDM) structure and galaxy formation have generally succeeded in
describing the observed photometric, stellar mass, size, star
formation, luminosity function, and clustering properties of LBGs,
especially by including varying amounts of dust extinction.  In
particular, the maximum stellar masses of LBGs predicted in smoothed
particle hydrodynamics (SPH) models are $M_{*, max} = 10^{11}$ at
$z=6$ and $10^{11.7}$ at $z=3$ \citep{nig06}; sizes are predicted to be
$r_{1/2} \sim 0.6-2 h^{-1}$ kpc and halo and internal velocity
dispersions to be
$\sigma_{halo} \sim 180-290~ \kms$ and $\sigma_{int} \sim 70-120~ \kms$,
respectively; star formation rates are expected to be $SFR \sim 15-100
\sfr$ \citep{mo99}; and SFR is predicted to correlate only weakly with
stellar mass \citep{wei02}.

The available information therefore suggests that LBGs at $z \sim 3$
range in stellar mass from $\sim 1-10\exp{10} \msun$ with total mass
from dynamical constraints $M<10^{11}\msun$, and that they reside in
dark matter halos with mass $M \sim 10^{12} \msun$, with significant
clustering.  Their star formation rates range from $10-100~ \sfr$, with
gas content and therefore star formation duty cycles and gas depletion
timescales poorly constrained.  Their comoving number density
correponds roughly to that of luminous galaxies today, and their
optical sizes and morphologies are diverse and compact.  The general
picture then is that LBGs are a heterogeneous set of actively
star-forming galaxies or sub-galactic clumps, building blocks that
will likely coalesce with other LBGs and/or non-LBG clumps to form
more massive galaxies, groups, and clusters by $z=0$.

To address further the open question of the dynamics of Lyman
break galaxies and their host halos and help place them with more
detail in the larger context of galaxy formation and evolution, we
have obtained spatially resolved optical (rest-UV) spectra with the
10-m Keck telescope of a sample of LBGs and searched for kinematic
information that we can use to constrain individual LBG and LBG host
halo dynamics directly.  Our sample is more than one magnitude fainter
and is at higher redshift than the kinematic targets previously
studied by others.

NIR IFU observations, especially with adaptive optics (AO)
\citep{for06, for09, leh09}, now provide direct access at $z \sim 2$ to \ha\ and other
rest-optical emission lines not strongly affected by dust, in contrast
to \lya.  However, optical spectroscopy such as that presented here
has the advantage of lower sky and telescope background emission and
higher sensitivity instruments and detectors, as well as broader redshift coverage.

The observations are described in \S~\ref{sec:obs}, evidence for
kinematic information is presented in \S~\ref{sec:kin}, and we discuss
the results in the light of current models, theory, and other
observations in \S~\ref{sec:disc}.  Throughout the paper we adopt the
currently favored cosmological parameters $\Omega_m = 0.3,
\Omega_{\Lambda}=0.7$, and $h = H_0 / 72~ \kmsmpc$.

\section{Observations and Data Reduction}\label{sec:obs}

\subsection{Lyman Break Galaxy Sample}

Our sample was drawn from 46 candidate and confirmed LBGs in the Hubble
Deep Field North (HDF-N).  We used the same sample selection as
\citet{low97}: blue in ``$B-I$'' ($B_{450} - I_{814}
<1.22$)\footnote[2]{Note that we are using AB magnitudes throughout,
transformed from ST magnitudes as detailed in the HDF information
posted on the World Wide Web: to transform from ST magnitudes to AB
magnitudes, add 1.31, 0.399, -0.199, and -0.819 to $U_{300},\
B_{450},\ V_{606},\ {\rm and\ } I_{814}$, respectively.} but extremely
red in ``$U-B$'' ($U_{300} - B_{450} > 1.41$), the colors expected for
a blue, star-forming galaxy at $2<z<3.5$ with a Lyman continuum break.
This color criterion is similar to that adopted by \citet{ste03}, but
is slightly less restrictive in the blue $U-B$, red $B-I$ region,
which \citet{low97} found allowed detection of "\lya-break galaxies" at
higher redshift in addition to pure Lyman break galaxies.  Seven of the 46 sources were included via that relaxed criterion. The galaxy magnitudes fell
in the range $ 24.01 < I_{814}< 26.27$, with a median value of 25.34.
The sample is thus more than 1 magnitude fainter than that of
\citet{pet01} and more than 2 magnitudes fainter than that of
\citet{erb06}.    We chose both targets with confirmed spectroscopic 
redshifts and those without, since the success rate for LBGs selected
in this way is typically so high ($> 90\%$ for bright sources,
$R<25.5$).   The median reported redshift was $z=2.7$.  Many of the galaxies discussed here were studied by
\citet{ste96b} and \citet{low97} but with lower spectral resolution
and/or less attention to the position angle of the spectrograph slit.

We took full advantage of the fine pixel scale and high resolution of
the HDF WFPC2 images, 0.04\arcsec pixel$^{-1}$ with $FWHM \sim$
0\farcs14.  We preferentially chose LBGs that showed clear signs of
one or more of the following: (1) significant spatial extent $>1$\arcsec\
in the HDF-N image; (2) multiple knots and/or close pairings of LBGs
in the HDF-N; and (3) previous spectroscopic evidence for some
kinematic features.  We also prioritized bright targets over faint
ones.  In all we observed 32 kinematic LBG targets in the HDF-N.  Of
the 32 total LBG targets observed, 21 showed some promise in the data
of allowing dynamical mass estimates; the remaining discussion will
focus exclusively on those 21 objects plus one serendipitously
discovered LBG, which together we call the ``high-priority targets''.
A complete list of all 22 targets is given in Table~\ref{tab:targs}, and the 11 unused targets are listed for completeness in Table~\ref{tab:unusedtargs}.  Most of those 11 were not useful for kinematic measurements because they had insufficient signal-to-noise ratio, no strong emission line, or no detectable spatial extent in our spectra.

We compared our target list to the list of LBGs hosting AGN in the HDF
according to {\it Chandra} X-ray flux as reported by \citet{lai06};
no sources match within 10\arcsec\ down to 0.5-2 keV flux levels as low as $3\exp{-17}\flux$ , so we conclude that none of our
targets hosts a luminous AGN.

\subsection{Observations}

Our observations were made during six observing runs from May 1997 to
April 1999 at the 10-meter Keck-II telescope using the Low Resolution
Imaging Spectrograph (LRIS; Oke 1995).  We designed six slit
masks around the LBG targets; high-priority targets were observed with
multiple masks to increase integration time.  Each target was examined
closely in the HDF image, and then its corresponding slit was tilted to
match the long axis of any extended emission from the target galaxy.
Slits were also tilted to cover multiple targets in close pairs or
groupings.  The masks were designed so that the position angle (PA) of
most slits was within $20^{\circ}$\ of the PA of the mask, meaning the
slits were aligned close to perpendicular to the dispersion direction.
But some slits were tilted as much as $40\degr$\ from the mask PA and
the normal to the dispersion direction.  Each mask had between 23 and
25 slits total (not including holes for alignment stars).  About half
the slits on each mask were designed around targets in the HDF and
Flanking Fields for other programs.  Slit lengths for our LBG targets
ranged from 8-35\arcsec, with a typical length of 15\arcsec.  All slits were
1\farcs1\ wide, although the slit tilts cause that width to be
projected to a dimension smaller on the sky by a factor of cos($i$),
where $i$ is the relative angle between the slit and the mask.  The
mask PAs were optimized for the anticipated mean hour angle of the HDF
at the time of observation, in the sense that we tried to keep the
majority of slits aligned with the parallactic angle to minimize
slit losses due to differential refraction.  We relied on our own
astrometric measurements of the HDF and Flanking Fields to derive
target positions.

For deriving wavelength solutions, HgNeArKr arc lamps were observed
through each mask.  Strong night sky lines were used to fine-tune the wavelength solution of some spectra. 

We used a 600 l/mm grating blazed at 5000~\AA, resulting in a
dispersion of 1.28~\AA~ pixel$^{-1}$, a typical resolution of 4 pixels
$\simeq$ 5~\AA\ full-width-at-half-maximum (FWHM) or 300 $\kms$.  This
represented a compromise between a lower-resolution grating that might
have delivered higher signal-to-noise per pixel and a
higher-resolution grating that would have provided more detailed
kinematic information.  The grating angle was adjusted so that central
wavelengths for slits at the mask centers were between 5000 and
6000\AA, giving a spectral coverage of 2620\AA\ and spectra ranging
from a minimum of 3456\AA\ to a maximum of 6978\AA.  The spatial pixel
scale for slits perpendicular to the dispersion direction was
0.215 arcsec pixel$^{-1}$, although this scale is compressed
(more arcsec pixel$^{-1}$) by 1/(cos $i$) for tilted slits, where $i$
is the relative PA between the slit and the mask.

Individual exposure times ranged from 1800 sec to 2700 sec, with at
least two exposures through each mask to help reject cosmic rays.
Spatial dithering was not possible due to the tilted slits.  Targets
received between two and 17 exposures each, depending on priority and
repeat placement on multiple masks.  This resulted in total exposure
times ranging from 9,900 to 50,000 sec per target.

The weather was mostly clear and photometric but some exposures were
affected by cirrus.   The seeing was typically 1\arcsec.

Images of all the targets are shown in Fig.~\ref{fig:images}, with the
location and PA of each slitlet superposed.  Targets observed at
multiple PAs are shown once for each PA.

We also included in our analysis some data from an earlier Keck/LRIS
observing run in April 1996 \citep{low97}.  The slits for that run's
slitmasks were tilted only to accommodate multiple objects, not to
align to the major axis of extended objects.  Nevertheless, some of
those slits did fall at the same PA as the slits in the observing runs
aimed at kinematic study, allowing us to add the datasets together.

The new observations are summarized in Table~\ref{tab:obs}, and properties
of the targets are listed in Table~\ref{tab:targs}.

\subsection{Data reduction}

We reduced the Keck/LRIS spectra using a combination of IRAF tasks and
custom IRAF scripts and C routines.  Raw images were
bias-subtracted, cosmic-ray cleaned, flat-field corrected, and
geometrically corrected to account for optical distortion.  

The next stage was carried out in one of two ways: (1) The wavelength
solutions for all the slits in a single exposure were measured
simultaneously, the image was rebinned so that all slits had the same
pixel-wavelength mapping and orientation perpendicular to the
dispersion, and then background sky emission was subtracted and
individual slits extracted as single two-dimensional spectra; or (2)
individual slits were extracted from each exposure and then treated as
long-slit spectra: a wavelength solution was derived from a
corresponding arc lamp image, the data were rectified, the wavelength
solution was fine-tuned using night-sky emission lines, and background
sky emission was subtracted.

Finally, for each slit all individual exposures were combined using
exposure-time weighting and spatial registration to make a single
two-dimensional, sky-subtracted spectrum.  One-dimensional (1D)
spectra of most targets were extracted from the two-dimensional (2D)
image.  Representative 2D and all the 1D spectra are shown in
Fig.~\ref{fig:2dA} and Fig.~\ref{fig:1da}, respectively.

\section{Redshifts and LBG Kinematics}\label{sec:mass}

In this section we present several new redshifts in the HDF, and then describe our measurements of kinematic features in the reduced spectra and how we use them to derive estimates of
dynamical masses, along with some significant caveats on those estimates.

\subsection{Redshifts}

We examined each one-dimensional spectrum visually to measure or
confirm the target's redshift.  For difficult cases we tried
cross-correlating the one-dimensional spectrum with various template
spectra, including an average of confirmed LBG spectra from
\citet{low97} and a high-signal-to-noise ratio (S/N) spectrum of the
gravitationally-lensed LBG MS 1512-cB-58 \citep{yee96, pet00} kindly
provided by M. Pettini.  In no case did the cross-correlation produce
a reliable new redshift.  In all, we found three tentative and one
robust new redshifts; confirmed, refined, or constrained four more;
and were unable to obtain useful redshift constraints for two.  We
were also able to measure the redshift for a new LBG the image of
which fell on one of the slits but that was not in our original target
list, for a total of five new redshifts.

Table~\ref{tab:zs} gives the redshift obtained for every object in the
sample.  The redshift quality $Q_z$ is given on a scale from 1 to 4
such that $Q_z=1$ means there is little hope of assigning a redshift
given the signal-to-noise ratio in our data, 2 means real features are
evident but the redshift is not secure, 3 means the redshift is
probable, and $Q_z=4$ means the redshift is definitely secure, with
multiple spectral features identified.

\subsection{Kinematic Measurements}\label{sec:techniques}

We next searched for evidence in our LBG spectra that would allow us to constrain their dynamical state or even estimate their masses or their host halo masses.   Measuring robust dynamical masses, however, is not straightforward, and the mass estimates we derive are subject to several major caveats.  We discuss some of the most significant of these below.

\subsubsection{Non-virialized systems: implications from simulations}\label{sec:nonvirial}

The velocities and velocity limits we measure may simply not
reflect orbital motion under gravity.  Peculiar motion of galaxies or
subgalactic clumps in unvirialized systems such as loose groups or very young clusters could
play a significant role.  Chance near superposition of physically unrelated sources could also lead to erroneous mass estimates.  Since many of our targets are indeed clumps of knots or close pairs, these factors may strongly affect our results and interpretation.

To try to assess the significance of the non-virialization scenario, we turn to the "phenomenological" n-body and semi-analytic $\Lambda$CDM galaxy formation models of \citet{kamp07}.   Their "Model 3" represents a combination of two scenarios: star formation dominated by quiescent star formation in disks, usually just the central galaxy, and star formation dominated by merger-induced starbursts (both galaxy-galaxy and halo-halo mergers).  One run of that model produced 13,257 simulated LBGs at $z\sim 3$ "detected" in 1 deg$^2$ with criteria similar to those used by \citet{ste03}.  We searched the catalog for projected close pairs of LBGs and examined their characteristics.  We find 1206 pairs with separation $r<7$\arcsec.  In Fig.~\ref{fig:deltav}, we plot the  relative velocity $\Delta v$ in each pair vs. the projected separation $\Delta r$ on the sky, where the velocities are calculated using the two galaxies' redshifts and their peculiar velocities.   We focus on the simulated pairs with redshift difference $\Delta z < 0.001$ (where "redshift" means cosmological redshift only, without considering peculiar velocity), which we take to be truly physically related.   There are 103 such pairs in the simulated catalog, or 8.5\% of all pairs with $r < 7$\arcsec; these are represented as open triangles in Fig.~\ref{fig:deltav}.   We also plot the values for the four close pairs in our sample (see \S~\ref{sec:disc}).  

 We calculate for each simulated close pair of LBGs a dynamical mass 

\begin{equation}
\label{eq:mdyn}
M_{\rm dyn} = r_{\rm dyn} v_{\rm rot}^2 / {\rm G}
\end{equation}

where $v_{\rm rot}$ is the maximum observed circular velocity, which in this case we take as $\Delta v$, and $r_{\rm dyn}$ is the radius at which the velocity is measured, in this case the separation between the pair.  We also calculate the total (dark matter) halo mass of each pair as the sum of the two individual halo masses listed for the galaxies in the simulation catalog.

Fig.~\ref{fig:mhmv} shows calculated dynamical mass vs. total halo mass for the close pairs in the simulation of \citet{kamp07}.  Surprisingly, there is no strong correlation seen.  This implies that the velocities in fact do not (or not always) reflect true virial velocities. They may instead reflect infall velocities of subclumps or galaxies streaming into a potential well; such infall velocities are likely larger than virial velocities by at least a factor of two.  We conclude that, if the simulation of \citet{kamp07} is correct, the pair-wise velocities in our LBG sample may provide at best weak constraints on the true total masses of our LBG systems.

\subsubsection{Incomplete spatial sampling}

A second caveat is that we may be sampling the kinematics of only a
small region, either central to a larger potential well or embedded
but not centered in a large massive proto-galactic cloud or even a
large but UV-faint galaxy.  
As discussed above in \S~\ref{sec:intro}, several recent
groups have used optical and NIR photometry to study the rest-optical
SEDs of LBGs in the HDF-N, and found that their stellar masses were
typically $M<10^{10}\msun$ \citep{pap01, dic04,saw98,rig06, lab05,
sha05,hua05}.  
These stellar masses provide a firm lower limit to the
total dynamical mass.

In the local universe, rotation
curves of massive disk galaxies rapidly reach maximum rotational
velocity $v_{\rm max}$, typically within a few kpc \citep{sof01},
similar to the half-light radii of the LBGs studied here.  Therefore,
if LBG kinematics behave at all like those of local disks then we
might not in fact be missing higher rotation velocities from faint
extended regions below our detection limit (although we could be
missing faint emission at the same velocity from larger radii, which
would raise the mass estimates).  Similarly, the central velocity
dispersions $\sigma$ of massive local elliptical galaxies are
generally excellent indicators of the total galactic virial mass as
measured by independent methods \citep{dez91}.  Again this implies
that even if LBGs reside in larger systems, the total mass may be
well-sampled by the LBG kinematics.  

Finally, we can compare LBGs at $z\sim3$ to luminous compact blue
galaxies (LCBGs) at redshifts $z<1$, which have many properties
similar to those of LBGs such as small half-light sizes $r_{1/2}<3
$kpc, high luminosities $L\sim L*$ (where $L*$ is the characteristic
luminosity in a Schechter luminosity function of galaxies today), blue
optical colors, diverse, irregular morphologies, asymmetries
$A\sim0.3$, star formation rates $10<SFR<100~ \sfr$, and narrow optical
emission lines, $\sigma < 100~ \kms$ \citep{guz96,phi97,pis01,guz03,low05}.  

\citet{bar06} found from deep optical
imaging that more than half of the 27 compact narrow emission line
galaxies (CNELGs, close cousins of LCBGs) at $0.1 < z < 0.7$ in their
sample had sizes consistent with small local dwarfs, arguing against
massive underlying host galaxies.  Similar results derive from studies
of nearby compact UV-luminous galaxies using GALEX
\citep{bas07,hec05,hoo07}.  The overall picture is that LCBGs at
intermediate redshift are galaxies with luminosity $L\sim L*$ but
masses only 1/10th that of a typical $L*$ galaxy, and
dynamical/stellar mass ratios $<3$.  

Constraints on stellar mass do not address the question of extended
massive dark matter or gas halos surrounding the LBGs, of course.
However, like  measurements of local galaxy halo masses using satellite
galaxy kinematics \citep{zar93}, our constraints on the dynamical
masses of close pair LBG systems are insensitive to mass-to-light-ratio or
stellar population uncertainties.

\subsubsection{\lya\ linewidth complications}

The third significant caveat regarding our mass estimates is that the \lya\ emission linewidths we use to estimate dynamical masses of individual LBGs and sub-clumps are subject to the influence
of bulk inflows and outflows of gas, strong absorption by dust, and
multiple scattering by neutral hydrogen, all of which can change the
line profile.  Such effects have been studied theoretically by several groups \citep[e.g.,][]{wol86, ver06, neu90, ten04}.   Observational evidence includes the systematic asymmetry and redshifting of the \lya\ emission line - presumably due to a combination of outflows and absorption by foreground dust --  in LBGs and local starbursts \citep{pet01, erb04, for06, erb06}, and \citet{ost08} find that the \lya\ emission of local starburst galaxies is spatially more extended than the continuum emission, presumably due to resonant scattering.  We  searched for but could not find any published or unpublished direct comparison between observed linewidths of \lya\ and those of \ha\ or other non-resonance nebular emission lines.  Therefore our mass estimates based on \lya\ emission line widths must be regarded as tentative.

\subsubsection{Unknown inclinations}

Finally, we generally lack constraints on the inclination
angle $i$ of our targets, so each mass has  associated with it an unknown sin $i$
correction factor.  Apart from estimating or
measuring inclinations of individual LBGs (as done by, e.g., Bouch\'e
et al 2007), this problem can be addressed only by assuming an average
inclination or else assembling a large enough sample of kinematic
measurements that all inclinations are statistically well represented.

\subsection{Mass Estimates}\label{sec:kin}

Given the caveats discussed above, we will proceed with caution in analyzing LBG dynamics from our sample.  Here we make an empirical distinction between {\em individual Lyman
break galaxies}, with any multiple knots of emission connected by
continuous flux visible in the HDF images and usually separated by
$\sim$1\arcsec\ or less, and {\em LBG systems}, consisting of two or more LBGs
without obvious extended emission bridging the gap(s) between them.
Both categories may correspond to star-forming galaxies or
sub-galactic clumps embedded in low-mass or massive dark matter halos,
and both may yield useful constraints on dynamical mass.  The LBG
systems offer an additional opportunity to measure the total host halo
mass as opposed to masses of subclumps or individual LBGs, although given the results of our examination of the simulation of \citet{kamp07}, we will not assume that such results are robust.

We examined visually the two-dimensional spectra of all 22 of the
high-priority targets for evidence of any of the following: (1) 
spatially extended emission lines, especially \lya\ but also including
\civ\ and \heii; (2) spectrally unresolved, resolved, or
multiple-peaked emission lines, especially from LBGs with multiple
knot morphologies; (3) extended continuum against which extended
absorption lines might be detected; and (4) emission and/or absorption
lines from close pairs of galaxies.  The features detected are listed
in Table~\ref{tab:zs}.

For each LBG or LBG system with evidence of kinematic information, we
then estimated the dynamical mass.  For simple gravitational rotation,
the dynamical mass would be $M_{\rm dyn}$ as in Eq.~\ref{eq:mdyn} above.

We divide our mass estimates into two categories: (1) masses estimated
from extended emission or absorption or multiple sources; and (2)
masses estimated solely from the \lya\ emission line velocity width
and observed half-light radius.  The measured velocity and size parameters and derived mass estimates for our sample
are listed in Table~\ref{tab:mass}. 

We compare in Fig.~\ref{fig:deltav} the distribution of velocities, which range from $<60$  to nearly 5000 $\kms$, and projected separations of the four close pair systems  with the prediction of Model 3 of \citet{kamp07}.  The pair with the largest $\Delta v$ appears to lie outside the locus of simulated pairs with small $\Delta z$ (triangle symbols; assumed to be physically associated, rather than just chance projection), but there are some simulated pairs with small $\Delta z$ and even larger $\Delta v$ than that target -- i.e., two LBGs within a few $\times 100$ kpc of each other but with very high relative velocity, $\Delta v \sim 10^5~ \kms$.  To assess the probability $P_{\rm true}$ that each pair in our sample is a true physical association, we can compare the density of small-$\Delta z$ pairs (triangles) to the density of large-$\Delta z$ pairs (dots) in different sections of the simulated $\Delta r - \Delta v$ plane.  We divide Fig.~\ref{fig:deltav} into eight equal segments bounded by 1, 4, and 7\arcsec\ and 0, 2000, 4000, 6000, and 8000 km/s.  Pairs with separation $\Delta r < 1$\arcsec are excluded because we would be unlikely to resolve such pairs spatially given the typical seeing in our data.  We calculate the probability $P_{\rm true} = N_{\rm true} / N_{\rm total}$,  where $N_{\rm true}$ and $ N_{\rm total}$ are the number of small-$\Delta z$ and total pairs, respectively, in each section.  We find that $P_{\rm true}$ ranges from 57\% for the two real LBG pairs ($\times$ symbol) in the lower left corner of Fig.~\ref{fig:deltav} to 6\% for the pair in the upper middle section.  In other words, according to the simulation, two of the four pairs have a better than 50\% chance of being true physical associations, while one has a 94\% probability of 
being a chance superposition.  The value of $P_{\rm true}$ for each pair is shown in Fig.~\ref{fig:m*}.

For each of the eight LBGs with \lya\ emission (six of which are also
included in the spatially extended emission category), we fit a
gaussian function to the emission line and measure the half-light
radius $r_{1/2}$ in the HDF $I_{814}$ image.  We do not deconvolve the
half-light radii with the WFPC2 PSF $FWHM=$0\farcs14.  This will have
only a small effect on the larger sources, but for the smaller sources
our measurement will overestimate the true half-light radius and
therefore the enclosed dynamical mass.  We did smooth the HDF images
with a gaussian filter in an attempt to constrain (and correct for)
the true seeing by matching the seeing-blurred spatial FWHM in the 2D
spectra.  However, significant uncertainties in the measured FWHMs
(especially for low S/N sources) and even variation within single 2D
spectra made this approach impractical.

We deconvolve each observed linewidth with the measured 5\AA\ spectral
resolution to derive intrinsic linewidths.  Based on the discussion
above of dynamical studies of local and high-redshift galaxies, we
adopt the FWHM of the line profile as a reasonable but untested proxy (modulo any
outflow and extinction effects) for circular velocity $v_{\rm rot}$,
we adopt $r_{1/2}$ for $r_{\rm dyn}$, and we calculate an estimate,
again based on Eq.~\ref{eq:mdyn}, of the dynamical mass within the
half-light radius.  These measurements and the linewidth-based masses
are also listed in Table~\ref{tab:mass}, and all the derived masses
are plotted vs. redshift in Fig.~\ref{fig:Mz}.  Six of the eight \lya\
emission lines were at best marginally resolved, so we plot the
derived dynamical masses for those sources as upper
limits.

\subsection{Notes on Individual Objects}

Here we discuss the results for each target, listed in order of
increasing RA, as in Table~\ref{tab:targs}.

\subsubsection{hd4\_0259\_1947}

This relatively bright, elongated source has a disky isophote and
captivated our interest as a potential high-redshift disk.  
\citet{coh00} report for their object No. 2 a redshift $z=0.904$, with
low confidence ($Quality=9$, where 11 is the lowest confidence).  Our
spectrum shows no sign of \mgii\ absorption at that redshift in either
2D or 1D, and no
\ciii\ (although at 3634\AA, our spectrum has low sensitivity).  We
conclude that the redshift assignment $z=0.904$ seems unlikely.

However, the final coadded spectrum shows weak emission lines at 4499,
5869, and 6045~\AA, plausibly matching \oii3727, \hb, and
\oiii5007, respectively, at redshift $z=0.207$.  We therefore
conclude (though with low confidence, given the lines' weakness) that
the source is a low-redshift interloper that slipped through our color
selection filter.

\subsubsection{hd4\_1076\_1847}

This compact, high-surface brightness source is the brightest target
in our survey.  \citet{low97} were unable to measure a redshift, but
found weak evidence for \mgii\ absorption at $z=1.0155$ and $z=0.879$.
\citet{coh00} report $z=0.882$ with low confidence
($Quality=9$, where 11 is the lowest confidence), and cite \citet{coh96}
as the redshift source, but we are unable to find the target in that
reference.  \citet{fer01} argued that the published spectrum
does not support that redshift, and added that their photometric
redshift technique favors $z_{\rm phot}=0.00$.  Meanwhile,
\citet{bud00} report a photometric redshift using HST/WFPC2+NICMOS
$z_{\rm phot} = 2.67$.

We accumulated 24.5 ks of integration time observing the source
with LRIS.  Nevertheless, we find no strong emission or absorption
lines, and we are unable to obtain an unambiguous redshift for the
source, nor can we measure any kinematic signature.

No strong continuum break is visible in our spectrum, which extends
down to 3320\AA\ with continuum detectable by eye in the 2D image down
to 3920\AA.  Since no strong \lya\ emission or absorption is seen
redward of that wavelength, we conclude that the emission redshift
must be $z<2.17$.

We again find weak evidence for \mgii\ absorption at $z=1.0155$ and
$z=0.879$, as reported in \citet{low97}.  Unfortunately the \mgii\
lines for the lower redshift coincide with night sky lines,
complicating their detection and measurement.  The lower redshift,
which is consistent with the $z=0.882$ value of \citet{coh00}, is
supported by possible \feii\ absorption.  If real, either or both sets
of redshifted absorption lines could be due either to intrinsic
absorption from the emitting galaxy itself, or to intervening
absorption if the emitting galaxy is at higher redshift.  The redshift
of the galaxy GOODS J123640.85+621203.4, with a position centroid
2\farcs2 from hd4\_1076\_1847, is reported by \citet{coh96} to be
$z=1.010$.  GOODS J123640.85+621203.4 has a highly linear morphology
extending at least 1\farcs5 in the HDF, aligned to within 10$\degr$ of
the separation vector between the two sources.  The close agreement
between its emission redshift and the tentative \mgii\ absorption
redshift we find in the spectrum of hd4\_1076\_1847\ suggests that it
is either responsible for the absorption, or else associated with
another as yet unidentified absorbing galaxy -- perhaps
hd4\_1076\_1847 itself.

We are therefore unable to confirm any of the previously published
redshifts or derive a new one for this source, apart from the lower
and upper limits mentioned above.  No kinematic information is
available from our spectrum.

\subsubsection{C4-09}

Object C4-09 in the catalog of \citet{ste96b} is a remarkable
source consisting of four bright knots of emission with nearly
identical colors, all within an area barely more than 1\arcsec\ across.
\citet{zep97} investigated and finally rejected the source as a
possible gravitational lens system.  As reported by \citet{ste96b},
the spectrum shows strong emission with two peaks at 5118.7 and
5132.6\AA.  Continuum is clearly detected redward of the emission but
not blueward, supporting the interpretation of the emission lines as
\lya\ at redshifts 3.211 and 3.222, respectively, slightly lower than
the $z=3.226$ reported by
\citet{ste96b}.  Weak absorption lines are seen at 5314, 5489,
5624, and 5633\AA, corresponding to \siii, \oi, \cii, and \siiv,
respectively at an average absorption redshift $z=3.216$, between the
redshifts of the emission line peaks.

We observed C4-09 with two PAs, one (98$\degr$) aligned with the
longest axis of the parallelogram of knots, and the other (137$\degr$)
closer to the short axis.  The four knots visible in the HDF image are
so close together, with spacings between 0\farcs57 and 1\farcs17,
that the continuum and the red side of the emission line appear
unresolved in both of our two-dimensional spectra.  The blue side of
the \lya\ emission line, however, appears to be slightly extended
spatially in the long-axis spectrum, with FWHM=1\farcs7, compared
with the seeing FWHM=1\farcs1, and offset towards the east with
respect to the redder line and the continuum.  No such spatial extent
or offset is visible in the short-axis spectrum.  No sign of ordered
rotation is visible.

The spatial extent and offset of the blue peak with respect to the red
peak and the continuum are consistent with a scenario in which some
subset -- one, two, or three knots -- of the four emission knots is
responsible for the blue peak, and another subset at a different
velocity gives rise to the red peak.  The blue and red \lya\ peaks are
well fit in the spectral direction by gaussian profiles with FWHM 5.5
and 7.0~\AA, respectively, i.e., unresolved or marginally resolved.
After deconvolution with the 5~\AA~ resolution, those widths
correspond to $< 135 \kms$ and $290 \kms$, respectively.  The 14~\AA\
separation between the two peaks, however, corresponds to a velocity
difference of 820 $\kms$.  If the wavelength separation is indeed due
to radial velocity differences among two or more of the four knots
(rather than, e.g., an absorption line superposed on a broad emission
line), then we can use it to estimate a system dynamical mass.  Given
the maximum spatial separation (which will yield maximum mass) of
1\farcs17, we derive a dynamical mass from Eq.~\ref{eq:mdyn} $M_{\rm
dyn} < 1.3\exp{12} \msun$.

Alternately, the double-peaked emission profile is also well-fit by a
single narrow absorption line with $\lambda = 5125.7$\AA\ and
$FWHM=8$\AA\ superposed on a broad emission line with $\lambda =
5126.5$\AA\ and $FWHM=14$\AA.  Such a broad emission line could be
caused by strong gas outflows, and the absorption line could be caused
by a foreground clump or screen of gas, even dust-free, that
resonantly scatters \lya\ photons out of the line center.  In this
scenario, the emission profile FHWM of 800 $\kms$ would reflect
outflow velocity, perhaps coupled with resonant scattering, rather than circular velocity of a virialized system.  The redshifts from the \lya\ emission and absorption lines would then
be $z=3.2164$ and 3.2170, in excellent agreement with the redshift
$z=3.216$ from the interstellar absorption lines but contrary to the commonly red-shifted \lya\ emission
reported by \citet{sha03} and \citet{erb04}, which they cite as evidence of outflows in LBGs.  Thus an outflow model for C4-09 appears somewhat problematic. 

We are not able with the current data to distinguish definitively between the two
scenarios.  If outflow is indeed responsible for the line
profile, then the true dynamical mass is most likely significantly
lower than the value derived above.

\subsubsection{hd4\_0818\_1037}

This source is compact and relatively bright, but continues to elude
efforts to obtain a definitive redshift.  

\citet{coh96} list a redshift $z=2.268$ and \citet{coh00} report
a corrected redshift $z=2.500$ with high confidence ($Quality=2$,
where 1=highest confidence).  However, we see no \alii 1671~\AA\
absorption matching either redshift, only possible \oi 1302~\AA/\siii
1304\AA\ and possible \nv 1243~\AA\ in P-Cygni profile matching
$z=2.500$; no \civ 1550~\AA\ is detected.  \citet{bud00} give a
photometric redshift $z_{\rm phot}=1.44$.

Continuum is clearly detected in our spectrum with a spatial profile
matching an unresolved PSF.  No strong emission or absorption lines
are detected.  Weak emission lines may be present at 5925 and 5976~\AA.
If these were \oiii\ at $z=0.194$, we would expect \hb\ at 5804~\AA\
and/or \oii\ at 4450~\AA, but neither is seen.  Possible weak
absorption lines appear at 4566, 4678, 5069, and 5447~\AA, but none is
well-detected, nor were we able to discern a pattern among them
indicating a source redshift.  We find no support for the tentative
redshift $z=2.04$ reported by \citet{low97} based on
cross-correlation, apart from the possible line at 5069~\AA, which
would correpond to \alii 1671~\AA.

Without solid absorption or emission line features available, we also
find no kinematic indicators in our spectrum.

\subsubsection{hd4\_0298\_0744}

hd4\_0298\_0744 is another elusive target.  \citet{coh00} report
$z=2.801$ with high confidence, while \citet{bud00} find $z_{\rm phot}
= 1.76$.  Continuum is clearly visible in the two-dimensional spectrum
even before sky subtraction down to the blue limit of our spectrum,
with no strong break visible.  No strong emission or absorption
features are seen in the final two- or one-dimensional extracted
spectrum.  We were therefore unable to confirm the redshift reported
by \citet{coh00}, which we would have expected to produce a strong
\lya\ emission or absorption line or continuum break visible at
4621\AA.  Because no strong continuum break is visible down to
3900\AA\ in our spectrum, we conclude that the redshift is most likely
$z<2.2$.

We cannot extract any dynamical information on the source.

\subsubsection{C4-06}

C4-06 from the catalog of Steidel et al. (1996) is one of the
brightest, largest sources in our sample.  The HDF image shows a
linear structure -- dubbed the ``Hot Dog Galaxy'' \citep{bun01} --
roughly $1\times3$\arcsec\ in extent, with two major knots and several
sub-knots.  The LRIS mask slit was aligned along that structure.

Strong, spatially resolved continuum is clearly detected in the
two-dimensional spectrum, although it appears as a single spatially
continuous source, rather than the separate knots seen with HST
resolution.  The spatial extent of the continuum emission, measured by
compressing the two-dimensional spectrum into a one-dimensional
spatial profile, is 3\farcs2, consistent with the HDF image.

No emission lines are visible, but several strong absorption lines are
detected matching redshifted \lya, \siii, \oi,
\cii, and \siiv.  The \lya\ line is broad (78\AA\ FWHM), while
the other lines are unresolved.  Additional marginal detections
include \siii, \ni, and \siiii.  The redshift derived from averaging
the strong narrow absorption line redshifts is $z=2.794$, while
cross-correlating the spectrum with that of a template average of 12
LBGs yields $z=2.802$, similar to the redshift $z=2.803$ reported by
\citet{ste96b}.

The narrow absorption lines show no sign of any velocity shift across
the extended continuum (see Figure~\ref{fig:2dA}).  To quantify any
subtle velocity gradient, we extracted three one-dimensional spectra
at the middle and the two ends of the linear source and intercompared
them.  The central extraction aperture was 12 pixels (2\farcs6) wide
and the apertures at the two ends were 6 pixels (1\farcs3) wide,
centered 8 pixels apart, so that the two end apertures had no overlap.

To constrain the velocity shift between the two end apertures, we
selected regions of the spectra containing strong absorption lines and
free from residual noise from strong sky emission lines and
cross-correlated them.  The cross-correlation yielded a strong peak at
$\delta \lambda = 0.8$\AA.  Assuming a minimum relative wavelength
accuracy of 1.0\AA, the velocity shift between the two apertures is
then $\delta v < 60~ \kms$, consistent with zero within our measurement
errors.  The dynamical mass limit implied by the observed constraint
on line-of-sight velocity gradient $\delta v < 60~ \kms$ over
3\farcs22 is $\mdyn < 2.6\exp9\msun$, the lowest dynamical mass
estimate in our sample (modulo, of course, the unknown inclination and
three-dimensional morphology of the source).

We interpret the lack of observed velocity gradient over more than 20
kpc as evidence that we are observing not an edge-on disk with ordered
rotation, but rather a truly linear, perhaps filamentary source.  The
object may of course be collapsing or even expanding perpendicular to
the line of sight; collapse is more natural under most current galaxy
formation scenarios.

\subsubsection{hd4\_1006\_0680}

This galaxy appears as a very close pair in the HDF image, with a
separation of only $0\farcs3$.  \citet{coh00} report a redshift for
the source $z=2.969$.  Our mask slitlet was aligned along the same PA,
$73\degr$, as the two emission knots; however, 1\arcsec\ ground-based
seeing certainly blurs the two into a single unresolved source in our
LRIS observations.

The spectrum shows very strong emission at 4823.2 and weaker emission
at 4812.3\AA.  The stronger line appears extended both spatially and
spectrally: the spatial FWHM of the line is 1\farcs2, comparable to
the seeing, but the emission profile's faint wings extend to cover
3\farcs3.  The spectral FWHM is 7.0\AA, compared to the resolution of
5\AA, and the total detected emission spans 26.4\AA, including the two
peaks and their wings.

The weaker line's spatial position centroid is offset 0\farcs25 from
that of the brighter line, closely matching the separation of the pair
in the HDF image.

Weak continuum is detected both redward and blueward of
the emission lines, but appears to be stronger on the red side.  No
other emission or absorption features are seen.

It is highly unlikely that this double emission line corresponds to an
optical line emitted by a low-redshift ($z<1$) system.  The \oii\
doublet would, at $z=0.293$, be separated by 3.5\AA, not the 11\AA\ we
observe.  Any \oiii\ or Balmer line should be accompanied by other
optical lines as well.  We conclude that the most likely
interpretation is \lya\ at $z=2.967$, confirming with only a slight
revision the redshift reported by \citet{coh00}.

The double emission line profile may be due to a single emission line
with strong absorption superposed by an intervening cloud or galaxy.
Alternatively, as for C4-09, it may be caused by radial velocity
differences between the two emitting knots.  Given the close match
between the spatial separations in the image and in the spectrum, we
adopt the latter scenario.  The velocity difference between the
emission peaks then corresponds to $677~ \kms$, while the entire
26.4~\AA\ span of the emission line corresponds to $\sim 1650 \kms$,
and the spectral FWHM of 7.0~\AA\ corresponds, after deconvolving with
the 5\AA\ instrumental resolution, to 4.9~\AA, or $\sim 300~ \kms$.

To derive a mass estimate for the system, we assume each knot in the
pair emits one emission line, and we therefore adopt $r=0\farcs3$
(2.3 kpc) and $\sigma = 677~ \kms$ to obtain $M = 2.4\exp{11}~\msun$.
The large total range in observed emission velocity, $1650~ \kms$, may
be due to bulk gas outflow induced by merging of two or more sub-units
and/or by the subsequent starburst.

\subsubsection{hd4\_1486\_0880}

Despite the relative brightness of this target, its redshift remains
elusive.  \citet{low97} reported a tentative redshift $z=2.47$ based
on a few possible absorption lines. \citet{tho03} estimated a
photometric redshift $z_{\rm phot}=2.80$, while \citet{bud00} give
none.  The source is smooth, slightly extended, and elongated in the
HDF image.  Continuum is easily detected along the entire length of
our spectrum.  However, no emission lines are visible, nor are any
strong absorption features.  There are possible absorption lines at
4386, 4528, 4794, 4816, 5168, and 5573\AA, but we were unable to
discern any convincing pattern matching our redshifted template
spectra or our previous tentative redshift $z=2.47$.
Cross-correlation with the template spectra likewise revealed no
robust redshift.  The lack of a strong continuum break redward of
4000\AA\ implies $z<2.29$, contrary to the photometric redshift
$z_{\rm phot}=2.80$ calculated by
\citet{tho03}.

With no strong absorption or emission features, we were also unable to
derive any kinematic information from the spectrum.

\subsubsection{hd4\_0367\_0266}

The HDF image shows the source hd4\_0367\_0266 to be an elongated
object with two bright knots separated by 0\farcs6 and an extended
tail terminating in a fainter knot, all nearly aligned.  The redshift
of $z=2.931$ reported by Lowenthal et al. (1997) was based on several
absorption lines.  The total extent of the source shown in the HDF is
2\farcs3; the LRIS mask slitlet was aligned to cover all three knots
as well as the tail.

Our spectrum shows weak extended continuum emission with FWHM $\sim$
3\farcs4.  No emission lines are detected.  Several possible weak
absorption lines appear in the extracted one-dimensional spectrum at
4956, 5117, and 5478\AA.  These lines would correspond to \siii, \oi,
and \siiv, respectively, at the previously reported redshift $z=2.931$.
A possible broad absorption trough at 4757\AA\ could be a damped \lya\
absorber at $z=2.913$.  Overall our confidence in the redshift is
somewhat lower than reported in Lowenthal et al. (1997), and we
downgrade the redshift quality to $Q_z = 3$.

To search for velocity gradients or shifts across the galaxy, we
extracted two spatially-independent one-dimensional spectra centered
2\farcs4 apart and cross-correlated them, using only the clean
regions of the spectra that were free from sky noise residuals and
showed evidence of absorption features.  The cross-correlation
function displays a weak peak consistent with $\delta v = 0$, but
unfortunately the signal-to-noise ratio was too low to allow any
robust measurement of kinematics.

\subsubsection{hd2\_1928\_1041 (C2-06) and C2-05}

The source hd2\_1928\_1041 (C2-06 in the catalog of Steidel et al 1996) is
separated by only 2\farcs0 spatially from
C2-05, and we consider them here as a close pair.  Each of the
galaxies is a compact source with one bright knot and a small cloud of
extended emission as seen in the HDF.  \citet{ste96b} reports $z=2.845$
for C2-05, but this is revised to $z=2.005$ by \citet{coh00}.  For
hd2\_1928\_1041, \citet{pap01} report $z=2.009$, citing \citet{low97};
we here slightly revise that redshift.

Both targets are covered in a single LRIS slitlet on our slit masks.
Strong, unresolved continuum is detected from both objects.  No
emission lines are visible, but numerous absorption lines
corresponding to interstellar species such as \siii, \oi, \cii, \siiv,
\civ, and \alii\ are well detected in the two spectra, corresponding
to redshifts $z=2.005$ (hd2\_1928\_1041) and $z=2.008$ (C2-05).  The
velocity difference implied by the redshifts is $\delta v = 390~ \kms$.
Cross-correlation of absorption line regions in the two
one-dimensional extracted spectra produces a strong peak at $\delta v
= 364 \pm 87~ \kms$ (where the error is estimated by dividing the FWHM
of the cross-correlation peak by 10), consistent with the simple
redshift difference.

The observed velocity difference and spatial separation provide a
dynamical mass estimate $\mdyn \sim 5.8\exp{11} \msun$ for the
two-source system.

\subsubsection{hd2\_1739\_1258}

A highly elongated, multiple-knot source embedded in extended
emission, extending over at least 2\farcs6 along the long
dimension, to which our LRIS slitlet was aligned.

Spectroscopic redshifts reported in the literature include $z_{\rm
spec}=2.72$ \citep{low97} with low confidence and $z_{\rm spec}=1.980$
\citep{coh00} with high confidence ($Quality=3$, where 1=best).
\citet{fer99} cite $z_{\rm spec}=2.002$, and \citet{cow04} cites
$z=1.98$, but we could not find the source of either measurement in
the literature.  \citet{wir04} reported that a redshift was not
measurable in their spectrum of the source.

Photometric redshifts calculated for hd2\_1739\_1258 include
$z_{\rm phot}=1.640$ \citep{fer99}, $z_{\rm phot}= 2.854$
\citep{tho03}, and $z_{\rm phot}= 1.34, 2.07$ \citep{bud00}. 

Strong, extended continuum emission is detected in our 39 ks
integration with a spatial $FWHM \sim $ 9 pixels, or 1\farcs9 and a
total extent of 3\farcs9.  The continuum dies below about 4075\AA,
implying $z\sim2.35$ if the dip is due to \lya\ blanketing.  However,
we also detect two possible weak absorption lines at 5305, 5318\AA,
which match \mgii\ at $z=0.897$.  These absorption lines could be due
to intervening gas, but they could also be intrinsic to the target.
We detect no features that support either the redshift $z=1.980$ 
\citep{coh00} or $z_{\rm spec}=2.72$ \citep{low97}.  
Given the lines we do detect weakly,
we tentatively conclude that hd2\_1739\_1258 in
fact lies at $z=0.897$ rather than at $z>2$.

We extracted two spatially independent 1-dimensional spectra from the
extended two-dimensional spectrum and cross-correlated them in an
attempt to measure velocity gradients across the source, but no
statistically significant cross-correlation signal was detected.

\subsubsection{hd2\_1398\_1164}

This galaxy consists of a single compact core with two short wings of
extended emission.  \citet{coh00} report $z=2.237$.  

Our LRIS slitlet was aligned with the long axis of the system.  The
spectrum shows a single weak emission line at 3934.2\AA~ with $FWHM
\sim 6$ \AA, i.e. barely resolved, with clear continuum redward but
none blueward.  Interpreting the line as \lya, which is supported by
the continuum break, we derive a redshift $z=2.236$.  No other strong
emission lines are detected, although possible weak \civ\ and \alii\
absorption and \heii\ emission are visible.  The width of the line
corresponds to a deconvolved velocity width of 240 $\kms$, which we
adopt as an upper limit.

Neither the continuum nor the strong emission line shows any spatial
extent in our two-dimensional spectrum, so we are unable to constrain
kinematics for this system.

\subsubsection{C3-02}

C3-02 is a small source with a single compact bright knot of emission
and a diffuse tail extending barely 1\arcsec, along which our LRIS slit
was roughly oriented.  \citet{ste96b} reported a tentative redshift
$z=2.775$,  \citet{fer99} report a photometric redshift $z_{\rm
phot} = 1.720$, and \citet{bud00} report $z_{\rm phot} = 2.218$. 

Spatially unresolved continuum is easily detected with average S/N$\sim11$ per resolution element over the entire
length of our 25.5 ks spectrum, which extends from 3650-6285\AA.  No
emission lines are visible.  Several possible weak absorption lines
appear at 4317, 4663, and 4608\AA, the first of those being the
strongest ($EW_{\rm obs} \sim 8$\AA).  None of those wavelengths
matches features expected from galaxies at $z=2.775$.  Both the
photometric and spectroscopic redshifts previously reported therefore
seem unlikely to be correct.  There is a possible match to \oi 1302,
\siii1304, \cii 1335/1336, and \siiv 1402/1403 at $z=2.316$.  No \lya\
emission is visible at the expected wavelength for that redshift, but
the continuum shape is consistent with a moderate break there.  Given
the weakness of the lines, we adopt this redshift only tentatively
($z_Q = 2$).  Kinematic measurements of this source are also
impossible from our spectrum.

\subsubsection{hd2\_0698\_1297}

hd2\_0698\_1297 consists of two distinct knots embedded in an
asymmetrical blob of diffuse emission.  The knots are 0\farcs5 apart,
and the diffuse emission visible in the HDF F814W image extends for
1\farcs2.  The system is 2\farcs7 from hd2\_0705\_1366, an unresolved point
source (measured $FWHM=$0\farcs14) in the HDF image.

Because of slit mask orientation constraints, we observed this pair at
two different position angles: for 25.5 ks at 42$\degr$, and for 10.2
ks at 58$\degr$.  Due to the close proximity of the two sources, the
1\farcs1-wide slitlets at both PAs include emission from both targets.

Both spectra show weak continuum with $S/N \sim 1$ per resolution element and a single emission line from each source.  The emission line from hd2\_0698\_1297 is stronger in the
spectrum with PA=42$\degr$, while that from hd2\_0705\_1366 is stronger
at PA=58$\degr$, presumably because of slit placement with respect to
each source.  We measure the wavelengths of the emission lines from
the stronger spectrum in each case: 5396\AA\ (hd2\_0698\_1297) and
5312 (hd2\_0705\_1366), each with $FWHM=8$\AA, corresponding for
redshifted \lya\ to $z=3.439$ and 3.370, respectively, within $\delta
z<0.01$ of the values reported in \citet{low97}.  There is evidence
for an absorption trough bluewards of the emission line in
hd2\_0698\_1297, supporting our identification of the line as \lya.
We cannot confidently rule out the possibility that each emission line
is in fact spectrally unresolved, so we adopt the measured FWHM as an
upper limit in each case.

The continuum and the emission lines are all spatially unresolved,
except perhaps continuum from hd2\_0698\_1297, which shows a slight
spatial extent $FWHM \sim$ 1\farcs5, although this may be due at least
partly to small spatial registration errors in co-adding the 18
individual exposures.  

Given the wavelength difference 84~\AA\ and projected separation
2\farcs7 between the two sources, and assuming the wavelength
difference to reflect gravitationally-induced dynamics, we derive a
system dynamical mass $M_{\rm dyn}=1\exp{14}\msun$, the largest
estimated dynamical mass in our sample.

\subsubsection{hd2\_0705\_1366}

     See hd2\_0698\_1297 (above).

\subsubsection{hd2\_0529\_1567}

See hd2\_0624\_1688 below.

\subsubsection{hd2\_0624\_1688}

This triple-knot compact source is part of a close pair with
hd2\_0529\_1567, 6\farcs25 away.  Both were covered by the slitlet at
$PA=28.9\degr$ that targeted hd2\_0725\_1818 and hd2\_0743\_1844 (see
below).  hd2\_0624\_1688, for which \citet{low97} reported $z=2.419$,
was in our original kinematic sample target list, while
hd2\_0529\_1567 was not: at $I_{814, AB}=27.233$, it offered little hope
of providing a useful spectrum.

However, our 2D spectrum shows clear continuum and a single strong
emission line from each source.  The continuum emission from
hd2\_0624\_1688 shows a significant drop blueward of the emission
line.  Continuum from the fainter hd2\_0529\_1567 is of course much
weaker, and the proximity of the emission line to the blue end of the
spectrum prevents us from assessing any continuum break across the
line.

The single emission line from hd2\_0624\_1688 lies at 4162.7\AA,
yielding redshift $z=2.424$ assuming that the emission line is \lya,
compared with $z=2.419$ reported by \citet{low97}.  No other
absorption or emission lines are visible, and \lya\ is the only
plausible interpretation.  The emission linewidth is $FWHM=6.3$~\AA,
corresponding to 3.8~\AA~ or $\sim 275~ \kms$ after deconvolution with
the 5~\AA~ resolution, and we adopt that value as an upper limit.

The emission line from hd2\_0529\_1567 lies at 3973.3\AA, yielding
$z=2.268$, again under the assumption that the line is redshifted
\lya.  The deconvolved linewidth is $FWHM=3.5$~\AA or $\sim 270~
\kms$, which we adopt as an upper limit.  No other absorption or
emission features are seen.

Together with the projected separation of 6\farcs25 (50 kpc), the
observed velocity difference of 1394 $\kms$ yields a dynamical mass
estimate for the pair system $M_{\rm dyn} = 2.2\exp{13}\msun$.

\subsubsection{hd2\_0725\_1818}

This small ($r_{1/2}=$ 0\farcs26) source lies only 1\farcs3 from
hd2\_0743\_1844.  \citet{low97} reported a spectroscopic redshift
$z=2.233$ for hd2\_0725\_1818 based on a \lya\ emission line, but only
a tentative redshift $z=2.39$ for hd2\_0743\_1844.  The two sources
have similar colors, with hd2\_0743\_1844 bluer by only 0.27 mag in
$B_{450}-I_{814}$ and redder by only 0.05 mag in $U_{300}-B_{450}$,
less than the spread in the colors of LBGs reported by \citet{pap05}; 
hd2\_0743\_1844 is fainter by 0.8 mag in $I_{814}$ (all measurements
from the catalog of Williams et al 1996).

Our LRIS spectrum covers not only both of those sources, but also a
spiral galaxy to the north with a redshift $z=1.148$ reported by
\citet{phi97} and, to the south, hd2\_0624\_1688 and hd2\_0529\_1567
(see above).

Continuum from both galaxies is easily visible in the two-dimensional
spectrum, but with some overlap of their spatial profiles, given their
mere 1\farcs3 separation.  The flux from hd2\_0725\_1818 is 2-3
times stronger than that hd2\_0743\_1844, consistent with their
photometric measurements.  Several absorption features are apparent in
the two-dimensional image; while they seem to span spatially the
continuum from both sources, the weakness of the fainter source makes
this appearance difficult to confirm visually.

We extracted a one-dimensional combined spectrum of the pair.  To
assess their redshifts and velocities independently, we also extracted
two separate one-dimensional spectra in the following way: First we
extracted a one-dimensional spectrum of the brighter source,
hd2\_0725\_1818, using only those pixels in the two-dimensional
spectrum extending spatially from the peak of the source away from
hd2\_0743\_1844, i.e. the side of the spatial profile that is less
contaminated by flux from the fainter source.  We then used this 1D
extraction to model the entire 2D spectrum and subtracted the scaled
model from the 2D image.  This left a clean, isolated spectrum of
hd2\_0743\_1844, which we then extracted to 1D.  We repeated the
process in reverse, modelling and subtracting the 2D spectrum of
hd2\_0743\_1844 to produce a clean, isolated and uncontaminated
spectrum of hd2\_0725\_1818.

The absorption lines in the 1D spectrum of hd2\_0725\_1818 are well
matched by interstellar SiII, CII, and AlII, and stellar SiIV, CIV at
$z=2.232$, thus confirming and slightly revising the redshift reported
by \citet{low97}.  We also find a tentative detection of \ciii\ in
emission, similar to their earlier result.  \lya\ lies at the blue end
of the spectrum, which is clean but low $S/N$; no emission or break is
visible in either 2D or 1D.

The spectrum of the fainter source, hd2\_0743\_1844, also shows
absorption corresponding to \siii, \civ, and \oi.  No \alii\ is
detected, though the $S/N$ is low in that part of the spectrum.  The
implied redshift is $z=2.230\pm0.004$, consistent with that of
hd2\_0725\_1818, and we adopt that redshift with quality $Q_z=3$
rather than 4 given the source's extreme faintness.  

To measure any small velocity difference between the two sources, we
cross-correlated the independent 1D spectra against each other, using
only those parts of the spectra free of bright sky emission line
residuals.  The result was a strong correlation peak (correlation
strength = 0.2) at $\delta v=9~\kms$, consistent with $\delta v = 0~
\kms$ within our nominal velocity resolution of $60~\kms$.  For
comparison, we also cross-correlated each 1D spectrum with a 1D
spectrum of blank sky extracted from the same 2D image.  No strong
correlation peak is seen anywhere near $\delta v=0$, with a maximum of
only 0.017 within $ 3000~\kms$ of $\delta v = 0$, and a maximum
correlation strength 0.1 for the whole spectrum.  We thus conclude
that the correlation between hd2\_0725\_1818 and hd2\_0743\_1844 is
significant, and that they have identical redshifts within our
uncertainties.

Adopting an upper limit $\delta v < 60~ \kms$ and the projected
separation 1\farcs3 or 10.9 $h^{-1}$ kpc, we then derive a dynamical
mass upper limit $M_{\rm dyn} < 8.6\exp{9}\msun$, the lowest derived
mass in our sample.

\subsubsection{hd2\_0743\_1844}

See hd2\_0725\_1818 above.

\subsubsection{hd3\_1633\_1909}

An elongated, wispy source consisting of two clumpy wings extending
over 2\farcs1, with a 0\farcs5 gap separating them, and aligned
roughly east-west.  The source's total isophotal colors in the HDF
Vesion 1 catalog satisfy the $B$-dropout criteria of \citet{low97}.
However, in the HDF Version 2 catalog, the two sides' colors are
significantly different, with the western side bluer by 1.30 mag in
$B_{450,AB}-V_{606,AB}$ and 0.52 mag in $V_{606,AB}-I_{814,AB}$.
Furthermore, in the Version 2 catalog photometry, neither component
individually satisfies either the $U$-dropout or the $B$-dropout
criterion.

Our LRIS slit covered both wings, as well as the source
hd3\_1824\_1945 7\farcs4 away and two other sources, one on either
end of the slit.  \citet{coh00} report $z=4.050$ for the source.

Only extremely very faint continuum is detected at the position of
hd3\_1633\_1909 in the 2D spectrum, with no absorption lines
discernible.  A single emission line appears, however, at 6145\AA.
The emission line, though apparently spectrally resolved with $FWHM =
8.4$\AA, is not well fit by the \oii\ doublet at $z=0.649$; the peaks
of the two components would be separated by 4.1\AA, while the line in
the spectrum is well fit by a single gaussian.  If the line is \lya,
the source's redshift is $z=4.056$, the highest redshift in our
sample and consistent with the redshift reported by \citet{coh00}.
The resolution-deconvolved velocity width of the emission line is
$FWHM=330 \kms$.

No spatially extended kinematic information is available from our spectrum.

\subsubsection{hd3\_1824\_1945}

This compact double-knot source was covered by the same slit that
included hd3\_1633\_1909 (see above), 7\farcs4 away.  \citet{coh00}
report $z_{\rm sp}=2.050$ (object 145 in their online list).  

Our spectrum shows strong continuum with spatial $FWHM=$1\farcs6
extending without any obvious break virtually to the blue limit of the
spectrum at 3750\AA.  No strong emission lines are visible; only a few
weak possible absorption lines are present.  We see no compelling
evidence to support the redshift $z=2.050$ reported by \citet{coh00};
for example, \alii 1671\AA, \civ, and \siiv, which often appear in
strong absorption in LBG spectra, are not seen.  

The lack of a strong \lya\ continuum break down to 3750\AA\ implies
$z<2.2$, in conflict with the photometric redshifts $z_{\rm
phot}=2.30$ calculated by \citet{fer01} and $z_{\rm phot} = 2.395$ by
\citet{bud00}.  Meanwhile, the lack of strong emission lines including
\oii\ 3727\AA\ implies $z>0.75$.

No kinematic information is available from our spectrum.

\subsection{Summary of results}

Of the 22 targets in our high-priority sample, we confirm that 14 are
LBGs with $z>2$ and $Q_z = 3$ or 4, of which 13 provided some evidence
of kinematics that we use to estimate dynamical mass or upper limits
thereon.  Two candidates are probably interlopers at $z<1$, and the
remaining six show insufficient features for confident redshift (or
kinematic) measurement.  The failure rate, 6/22 = 27\% (or 8/22=36\%
if we include the two interlopers), is higher than the typical failure
rate for LBGs with $R<25.5$, $<10\%$.  This must be due to our color
selection function, which is somewhat different from that of
\citet{ste03} and may be less restrictive, as mentioned above; our
magnitude limit, which is 1-2 mags fainter than most previous
spectroscopic surveys of LBGs; our use of the 600~l/mm grating, which
provides less spectral coverage than the lower-resolution gratings
commonly used for LBG redshift surveys; or a combination of those
effects.

Eleven of the 13 LBGs with kinematic signatures show spatially
extended emission or absorption: eight are in close pairs, two contain
three or four knots of emission each, and one is a clumpy elongated
source extending over 3\arcsec or $\sim$ 25 kpc.  A different subset
comprising eight of the 13 are sources with \lya\ emission lines whose
linewidths may provide some constraint on dynamical mass.  The redshifts
of the combined subsets range from $z=2.005$ to $z=4.056$ with a
median of $z=2.424$.  The angular extent of emission ranges from
0.12-6\arcsec ($\sim 1-50 h^{-1}$ kpc), and the adopted circular
velocities range from $< 60~\kms$ to 4700~$\kms$, with a median of 330~$\kms$.

We divide the mass estimates into two overlapping categories based
on the source morphologies and angular extent: individual LBGs,
comprising 11 sources, and LBG systems, comprising the four close
pairs.  We derive dynamical masses for the LBGs (using \lya\
linewidths and extended and multiple-knot sources) ranging from
$<4.3\exp{9}\msun$ to $1.3\exp{12}\msun$, with a median $M_{\rm dyn} =
4.2\exp{10}\msun$.  We see no evidence in any of our targets for
ordered rotation such as measured by \citet{for06} and \citet{erb06}
for some of their more luminous star-forming galaxies at lower
redshift ($z \sim 2$).  Dynamical mass estimates for the LBG systems range
from $<10^{10}\msun$ to $\sim 10^{14}\msun$, with a median mass
$M=10^{13}\msun$.  We emphasize that all these values are subject to the caveats discussed in \S~\ref{sec:techniques}, and that two of the four close pairs have probability $> 70$\% of being chance projection rather than physical associations, according to our comparison with the models of \citet{kamp07}. 

The tentative dynamical masses we derive for individual LBGs, $<4 \exp{9} h^{-1}
\msun $ to $ 1.1 \exp{11} h^{-1} \msun$, are very similar to those
measured for brighter targets by \citet{for06} based on \ha\
kinematics measured with VLT/SINFONI, $m_{\rm dyn} \sim
0.5-25\exp{10}\msun$, implying that either \lya\ and \ha\ emission
linewidth measurements are both robust or else they both suffer from
complications of comparable scale.
\citet{for06} also extrapolate the velocities beyond the half-light
radii to estimate the virial mass of a typical halo, finding $m_{\rm
halo} \sim 10^{11.7\pm 0.5} \msun$, similar to the halo masses
inferred by \citet{ade05a} based on clustering.  In contrast, the
median dynamical mass $1.1 \exp{13}\msun$ that we measure for our four
close-pair systems is an order of magnitude larger, and the maximum
system dynamical mass estimate is $>10^{14} h^{-1} \msun$, more than
$5\sigma$ above the mean halo mass of \citet{for06}.  
We note, however, that in no case in our sample is the derived dynamical mass less than the reported stellar mass for the same galaxy (\S~\ref{sec:mstar}), which is reassuring.

\section{Discussion}
\label{sec:disc}

Having measured redshifts and searched our spectra for dynamical signatures 
in 13 LBGs and LBG systems, we now compare those results to several other observational results and to theoretical expectations in order to 
explore implications for galaxy formation scenarios.

\subsection{Mergers, close pairs, and implications for galaxy formation}

Many lines of evidence point to the importance of merging in galaxy
evolution over the entire cosmic history of galaxies.  In this section we explore the possibility that our sample of LBGs includes a significant merger fraction, and we compare the characteristics of our sample directly to the predictions of galaxy formation theories, including masses and close pair fractions.  

Cold accretion of gas and dark matter not associated with major majors is also increasingly appreciated as a significant source of fresh material for star formation and mass buildup \citep{gen08 , coo09, age09}.  
But major mergers, where the components have roughly equal mass, can erase disk
morphologies and produce starbursts, AGNs, and elliptical galaxies,
and even minor mergers can induce bursts of star formation or
disturbed morphologies \citep{bar92,mih96, lin07}.    Recent works have measured the evolution of the rate of galaxy mergers, the fraction of close pairs of galaxies, and the distribution of galaxy morphologies over cosmic time \citep{kam07, bri07, lin04, bel06, kar07, lot06 , con03}.  Essentially all of those results are consistent with hierarchical galaxy formation models, in which galaxy
mass is assembled largely through merging.  

As noted above, early studies of LBGs recognized the high rate of
close pairs and multiple knots \citep{col96,col97,low97}.  Of the 14
confirmed LBGs in our sample, 12 are in close pairs or have
multiple-knot or highly extended, clumpy morphologies.  All 12 are
therefore excellent candidates for merging systems.  Indeed, every
close pair in our sample yielded potential kinematic information;
conversely, the targets that failed to provide such information were
predominantly low-surface brightness, elongated, diffuse sources,
which may themselves be interacting or merging systems.  Our entire
sample, of course, is strongly biased away from isolated, symmetrical
morphologies by virtue of our visual selection process, so it is to be
expected that the sample would show a higher fraction of merging
systems than do magnitude-limited LBG samples.

Our sample contains only four close pair systems, so it
is subject both to small number statistics and especially to the
assumption that the pairwise velocities indicate dynamical mass (see \S~\ref{sec:techniques}).  The
components of the four pairs may be subclumps in various stages of
infall and virialization, in which case their relative velocities could
be dominated by random rather than circular motions.  They may,
however, also reflect a true, large range in the masses of the LBG
host halos.

The theoretical implications of observed LBG close pair incidence and
velocity distributions for galaxy formation models have been explored
in detail by several groups \citep{som01, wec01,zha02,zha03a,zha03b,shu01}.
\citet{wec01} found that different models predicted very different halo occupation
distributions, i.e. the number of LBGs per dark matter halo.  In most
$\Lambda$CDM models, the number of LBGs predicted per halo increases with
mass, with typically $\sim 1$ LBG halo$^{-1}$ for $M
\sim 10^{12} \msun$ and $R<25.5$.  The number of close pairs of
galaxies -- defined as galaxies within a given angular separation
within a redshift interval $\Delta z < 0.04$ -- is an especially
sensitive discriminator between models.  Specifically, compared to the
observed number of close LBG pairs, models postulating a one-to-one
correspondence between LBGs and massive dark matter halos (``massive
halo'' models) underpredict close pairs by $1.5 \sigma$, while the
``colliding halo'' model overpredicts close pairs by $4 \sigma$ within
15\arcsec.  The best agreement was provided by their ``collisional
starburst models'', in which LBGs host both ongoing quiescent star
formation and rapid bursts of star formation triggered by minor and
major galaxy mergers.  

To compare the observed incidence of close pairs in our sample with
the models of \citet{wec01}, we first note that 
only two of the four systems we call ``close pairs'' satisfy the
\citet{wec01} criterion $\Delta z < 0.04$.  Fig.~6 in that paper shows
the fraction $N_{\rm pairs} / N_{\rm galaxies}$ as a function of pair
separation for different models.  All four of the pairs in our sample
have separation $<15$\arcsec, so they all fall in the first bin in the
figure.  If we assume that we have observed all the close pairs in the
HDF survey volume in that separation bin, i.e. the total number of
close pairs with $\Delta z < 0.04$ is $N_{\rm pairs} = 2$, and further
that the total number of galaxies in the survey volume with comparable
selection criteria is roughly the number of LBG candidates in the HDF
times the success rate we observe here, or $N_{\rm galaxies} \sim 46
\times 64\% = 29$, then we find $N_{\rm pairs} / N_{\rm galaxies} \sim
0.07$.  This value is already marginally inconsistent with (higher
than) the values, all $\sim 0.04$, predicted by the collisional starburst, constant
efficiency quiescent, and accelerated quiescent models in Fig.~6 of
\citet{wec01}; the massive halo model underpredicts close pairs by a
factor of more than $10\sigma$ (including only the error bars of
Wechsler et al 2001).  The colliding halo model, on the other hand,
overpredicts it by a factor of less than $1\sigma$.  Obviously, with
such small numbers of close pairs and large uncertainties on the
selection function and completeness of our sample, the observed
incidence ratios should be viewed as only tentative.

The close pairs and multiple knot systems we study here all have
maximum angular separation $r<7$\arcsec, and therefore all fall in the
first bin in the separation-pair fraction plot (Fig. 6) of
\citet{wec01}, where indeed the discrimination among models is the
strongest.  Because of the biased nature of our target selection, we
do not attempt to compare quantitatively the observed pair fraction
with the predicted one.  But we do note that close pairs are
represented over the entire range of dynamical masses we derive,
$10^{10} - 10^{14} \msun$.  This distribution is likely due either to (1)
misinterpretation of the kinematic signatures (i.e. the velocities are
not due primarily to orbital motion under gravity), resulting in
erroneous mass estimates or (2) misidentification of individual knots
of emission as separate galaxies, thus complicating the comparison
with the models; or else (3) it is inconsistent with the general
prediction of \citet{wec01} that the number of LBGs per halo should
increase with halo mass.

\citet{shu01} use observed sizes and SFRs of LBGs to conclude from
their models that LBGs must be hosted by halos with circular
velocities $v_c = 100-300~ \kms$, although for some combinations of
gas-to-total and dust-to-total mass ratios the distribution can reach
400 $\kms$.   If LBG close-pair dynamics are indeed indicative of host halo
mass, then the range of halo masses in our sample is significantly larger
and the typical mass significantly higher than predicted by
\citet{shu01}.  \citet{zha02} further explore the possibility of using the observed
pairwise velocity distribution (PVD) to discriminate between different
galaxy formation models, but their investigation focuses on scales between 100 $h^{-1}$ kpc and 30
$h^{-1}$ Mpc, much larger than the 10-kpc scale of multiple-knot and
close pair targets studied here.

\citet{zha03a} analyze dark matter halo $N$-body simulations and 
report that mass accretion occurs in a rapid phase, during
which the halo's circular velocity is also predicted to rise rapidly,
followed by a slow phase, during which the circular velocity remains
almost constant.  However, they examine only isolated halos with total
mass fixed at $M=10^{12} h^{-1} \msun$, for which the maximum circular
velocity predicted is $\sim 300~ \kms$ and the maximum halo radius is
$\sim 30 h^{-1}$ kpc.  These spatial scales are roughly comparable to
those of three systems in our sample: two close pairs and the highly
extended source C4-06.  The velocity scale is comparable in
only one close pair system.  Most of the LBGs and LBG systems in our
sample are therefore not apparently well described by the models of
\citet{zha03a}.  However, if we interpret our results within their
framework, the large range in system dynamical masses we derive could
imply that we are seeing the systems at different evolutionary stages,
some before the rapid mass accretion phase is completed and the
circular velocity plateaus, and some after.  Typical group-sized halos
today have circular velocities $\sim 500~ \kms$, and cluster halos have
circular velocities $\sim 1000~ \kms$.  Only one of our LBG system
velocities lies well above that range, with $\Delta v = 4700~ \kms$;
the rest have velocities consistent with those of normal galaxies to
normal clusters of galaxies today.  As noted above in \S~\ref{sec:kin} (Fig.~\ref{fig:deltav}), the distribution of pairwise velocities and projected separations of our four pairs is consistent with the predictions of Model 3 of \citet{kamp07}, supporting their scenario of quiescent disk star formation combined with merger-induced starbursts.

The dynamical masses we derive span the range of masses of dwarf
galaxies to giant elliptical galaxies today.  Our modest sample is
certainly biased and is not representative of all LBGs at $z\sim 3$.  The range of estimated dynamical masses is similar to that derived for simulated LBGs in Model 3 of \citet{kamp07} (\S~\ref{sec:kin}), but the kinematics of both simulated and real datasets may in fact not reflect true virial masses, especially given the lack of correlation with combined halo mass in the simulation (Fig.~\ref{fig:mhmv}).
However, taken together, the sizes, SFRs, morphologies, velocities,
and derived individual and system dynamical masses of the LBGs in our
sample imply a large range of physical scenarios and dark matter halo
masses.  They also further suggest that, rather than being identified
one-to-one with the centers of future massive spheroids, bulges, or
elliptical galaxies, LBGs are merging sub-clumps within halos that are
destined to become galaxies with a large range of masses and
potentially of morphological type as well.  This scenario is
consistent with hybrid galaxy formation models such as the favored
collisional starburst model of \citet{som01} and \citet{wec01}, but
not with their massive halo model.

\subsection{Kinematic vs. Stellar Masses}
\label{sec:mstar}

Here we compare our kinematic mass estimates to the stellar masses of the same
galaxies calculated by SED fitting.

\citet{pap05} expanded on the sample of \citet{pap01}, fitting stellar
population models to SEDs of UV-bright galaxies (i.e. LBGs) at
$1.9<z<3$ in the HDF based on fluxes from HST/WFPC2 and NICMOS.  The
stellar masses for the 53 galaxies with $S/N>20$ in the NICMOS F160W
band range from $1\exp{8} - 3\exp{10} \msun$, with a median around
$3\exp{9} \msun$.  The stellar mass estimates for 765 of 1682
NICMOS-detected sources with spectroscopic or photometric redshifts
have since been supplemented by photometry from {\it Spitzer}/IRAC and
were kindly provided to us by C. Papovich (2007, private
communication).

We searched the {\it Spitzer}-supplemented NICMOS+WFP2 catalog for
positional matches to our high-priority LBG sample.  All 22 targets
have at least one match well-detected by NICMOS within 0\farcs6.  We
visually examined and confirmed the position of each match in the HDF
F814W images.

Of the 22 LBG targets, 20 have stellar mass estimates in the Papovich
catalog, the other two lacking reliable redshifts.  Neither of those
two, however, is among the 13 LBGs in our sample showing kinematic
signatures.  Therefore we can directly compare all the individual and
system (i.e., halo) dynamical masses we estimated to the stellar
masses estimated by \citet{pap05}.  The  HDF Nicmos Mosaic (HNM)
catalog numbers, redshifts used, and stellar masses of all matched
objects are listed in Table~\ref{tab:m*}.

Six targets have two matches in the Papovich catalog within 1\farcs3.
Of those, three are in our list of 13 with dynamical mass estimates:
C4-09, C4-06, and hd3\_1633\_1909, each producing two matches.  Both
matches in each case correspond to emission knots we consider to be
part of the same LBG.  For C4-09 and C4-06, the redshifts used by
\citet{pap05} to calculate SED fits and stellar masses were for both
matches essentially the same redshifts we report here.  For
hd3\_1633\_1909, however, \citet{pap05} used the same redshift we find
($z=4.056$) for only one of the two matches, while for the other they
used $z=2.175$, i.e. what we consider to be a single LBG they consider
to be two sources at widely disparate redshifts.  The resulting
stellar mass estimate for the low-redshift source (HNM 1499) is $\sim$
1/3 that of the high-redshift source (HNM 1506).  We consider below
the stellar mass estimate of only the high-redshift source.  The
low-redshift source certainly contributes flux to our spectrum.
However, as our dynamical mass estimate of hd3\_1633\_1909 is based
only on the \lya\ emission line width, the effect on our mass estimate
should be negligible.  On the other hand, the redshift $z=2.175$ of
the low-redshift source may be in error; if the correct redshift is in
fact close to $z=4.05$, then the stellar mass estimate for
hd3\_1633\_1909 would increase by roughly a factor of two.

We make the following three comparisons of the stellar mass estimates to our
dynamical mass estimates: (1) for the eight \lya\ linewidth-based
dynamical mass estimates or upper limits, we compare those values
directly with the single matching stellar mass estimates (seven
sources) or the two components of a double match (one source, C4-09)
for a total of nine comparisons; (2) for the three individual but
extended or clumpy sources with single or multiple matches in the
NICMOS catalog, we combine the stellar mass estimates and compare the
total to our individual LBG dynamical mass estimated from extended
absorption or emission velocity or, if neither of those is available,
from \lya\ linewidth; and (3) for the close pairs, we combine the two
stellar mass estimates and compare the sum to the system dynamical
(halo) mass from the pair's relative velocity.  The four-knot source
C4-09, which yields both
\lya\ linewidths and a double emission line peak, presumably from
different emission knots, has two NICMOS matches, so we apply both
comparison types 1 and 2 (i.e. we separately compare the smaller
dynamical and stellar masses -- perhaps due to individual knots -- and
the combined stellar and system dynamical masses).  These comparisons
are shown in Fig.~\ref{fig:m*}.

The individual LBG stellar masses (comparison types 1 and 2 above)
range from $2.50\exp{8}$ to $1.6\exp{10} h^{-1} \msun $ with a median
of $1.6\exp{9}\msun$.  Two sources show stellar masses roughly equal
to their dynamical masses.  The stellar mass range and median,
however, are both less than the range and median of dynamical masses;
the typical mass ratio $M_{\rm dyn}/M_* \sim 5$, comparable to those
of LCBGs at $z<1$ \citep{guz03}.  The stellar masses are also small
compared to those measured for 181 LBGs in the EGS by
\citet{rig06},  $10^{9-12} \msun$ with median $2.95\exp{10}\msun$.
No obvious correlation is seen in Fig.~\ref{fig:m*} between our
dynamical mass estimates and the stellar masses from \citet{pap05},
although a trend could be masked by the six upper limits on dynamical mass
from \lya\ linewidths.

\citet{erb06} likewise compared stellar and dynamical masses for 68
galaxies at lower redshift $z\sim2$, finding a weak correlation
between the two, with a mean stellar mass $M_*=3.7\exp{10} \msun$ --
more than 20 times the median stellar mass for our fainter,
higher-redshift sample from \citet{pap05} -- and a typical mass ratio
$M_{\rm dyn}/M_* \sim 2$ vs. 5 for our sample.  They also found that
$\sim 15\%$ of their sample had large dynamical-to-stellar mass ratios
$M_{\rm dyn}/M_* > 10$ (vs. 10/12 = 83\% for our sample), suggesting
that this was likely due to young galaxy ages that had not yet built
up significant stellar mass, or else faint older stellar populations
that escaped detection.

The same scenarios may explain the large dynamical-to-stellar mass
ratios and lack of correlation we observe.  Our sample is more than
one magnitude fainter than that of \citet{erb06}, so we are probing
farther down the LBG luminosity function.  Local galaxies and even
starburst galaxies show a wide range in mass-to-light ratio, from $M/L
\sim 0.01$ in solar units for extreme starbursts such as \hii\
galaxies to $M/L \sim 10$ for quiescently star-forming galaxies like
the Milky Way \citep{guz96}.  The large range is presumably due
to a combination of galaxy age (older systems showing stronger NIR
flux from stars), burst age (younger systems showing stronger UV flux
from stars), merging history, and dust obscuration.  The large range
we observe in dynamical-to-stellar mass ratio may in fact reflect a
wide range of intrinsic LBG properties including dark matter halo or
sub-clump mass, star formation history, merging history and status,
and burst age.

We also compared the system masses derived from the four close pair
relative velocities with the combined stellar mass estimates from
Papovich et al (2005; comparison type 3 above).  The right-hand panel
of Fig.~\ref{fig:m*} shows the comparison.  Here the range in combined
stellar mass is smaller than for the individual LBGs, while the range
in system dynamical mass is larger.  The average total stellar mass is
$5.0\exp{9} \msun$, and the average system dynamical-to-stellar mass
ratio is $1.1\exp{4}$, ranging from $M_{\rm dyn}/M_* < 3$ to $M_{\rm
dyn}/M_* = 2.6\exp{4}$.  This large mass ratio range suggests that one
or both of two scenarios are at play: (1) the LBGs are embedded in
dark matter halos with a wide range of masses that are not strongly
correlated to the luminosity and star formation rates of the LBGs they
host; and (2) the relative velocities of the close pairs may reflect
various stages of infall, i.e. not all the close pairs are necessarily
in virialized systems (see \S~\ref{sec:techniques}).  

\section{Summary and Conclusions}
\label{sec:summ}

We have studied candidate and confirmed LBGs at $z \sim 3$ in the
HDF-N with deep two-dimensional spectroscopy, finding 3 tentative and 2 robust new redshifts.  We also find evidence for
kinematics in 13 LBGs and four LBG close-pair systems.  The observed morphologies and dynamics, including numerous cases of close pairs and multiple sources, appear to be most consistent with collisional starburst models of galaxy formation, rather than massive halo models with only one LBG per halo.  We derived tentative
dynamical mass estimates from the observed velocity signatures, finding a
range from $4.3\exp{9}\msun$ to $1.3\exp{12}\msun$ with a median
$M_{\rm dyn} = 4.2\exp{10}\msun$ for the individual LBGs and a range
from $<10^{10}\msun$ to $\sim 10^{14}\msun$ with a median mass
$M=10^{13}\msun$ for the close-pair systems.  All the mass estimates are subject to strong caveats including unknown inclination corrections, \lya\ absorption, outflows, and
non-virialized systems.  
All our dynamical mass estimates are larger than the corresponding stellar mass measurements.
However, we find no evidence for a correlation between stellar mass and
dynamical mass, in contrast to the weak correlation found by
\citet{erb06} for their sample of brighter, lower-redshift sources.

We argue on the basis of comparison with local compact starburst galaxies, other studies of distant star-forming galaxies, and galaxy formation theories and simulations that the sizes, SFRs, morphologies, velocities, and dynamical mass estimates imply a wide range of physical scenarios and mass scales giving rise to LBGs at $z \sim 3$.

Given the large time investment at Keck needed to observe even this
small sample, further progress in constraining the dynamical masses of
LBGs will require technological advances including the deep
diffraction-limited imaging and spectroscopy promised by {\it James
Webb Space Telescope} and increasingly available through adaptive
optics and NIR integral field spectroscopy at ground-based 10-meter
class telescopes.

\acknowledgments

Many thanks are due to the staff of the Keck Observatory for their
help and expertise in carrying out the observations.  We appreciate
assistance from Sandra Faber, Drew Phillips, and Nicole Vogt during the observations, and from Dan Kelson, Gabriela Mallen-Ornelas, and Nathan Roche
in reducing the data.  Thanks to M. Pettini for providing the
comparison spectrum of cB-58, to C. Papovich for the electronic
catalog of LBG stellar masses, and to D. Hogg for help with the color image
recipe.   J.D.L. acknowledges support from NSF grant AST-0206016.  This
research has made use of NASA's Astrophysics Data System Bibliographic
Services; the NASA/IPAC Extragalactic Database (NED) which is operated
by the Jet Propulsion Laboratory, California Institute of Technology,
under contract with NASA; and IRAF.

Facilities: \facility{Keck(LRIS)}, \facility{HST(WFPC2)}.

%%%%%%%%%%%%%%%%%%%%%%
%
%   REFERENCES
%
%%%%%%%%%%%%%%%%%%%%%%

\clearpage

%%%%%%%%%%%%%%%%%%%
%
%    TABLES
%
%%%%%%%%%%%%%%%%%%%

%%%%%%%%%%%%%%%%%%%%%%%%

\begin{deluxetable}{lrrrrrrrrl}
\tablecaption{Observing Runs at Keck II Telescope \label{tab:obs}}
\tablewidth{0pt}
\tablehead{ 
\colhead{Date (UT)}
}
\startdata
1997 May 5-6 \\
1997 May 30  - June 1   \\
1997 June 6  \\
1998 April 28 -- May 1 \\
1998 May 23-25 \\
1999 April 15-18 \\

\enddata
\tablecomments{The Low Resolution Imaging Spectrograph with a 600 l
mm$^{-1}$ grating blazed at 5000\AA\ and slit widths of 1\farcs1 were
used for all observing runs, giving spectral resolution about 300
$\kms$ FWHM (130 $\kms$ gaussian $\sigma$).}
\end{deluxetable}

%%%%%%%%%%%%%%%%%%%%%%%%%%%%

\begin{deluxetable}{lrrrrcl}
\tabletypesize{\tiny}
%\rotate
\tablecaption{Lyman Break Galaxy Kinematic Targets in the {\em Hubble Deep Field.}\label{tab:targs}}
\tablewidth{0pt}
\tablehead{ 
   \colhead{Name\tablenotemark{a}} & 
   \colhead{HDFv2 \tablenotemark{b}} & 
   \colhead{RA (2000)\tablenotemark{c}} &
   \colhead{DEC (2000)\tablenotemark{c}} & 
   \colhead{$I_{814,AB}$\tablenotemark{d}} & 
   \colhead{PA\tablenotemark{e}} & 
   \colhead{$t_{\rm int}$ \tablenotemark{f}} \\

   \colhead{~} & 
   \colhead{~} & 
   \colhead{~} &
   \colhead{~} & 
   \colhead{mag} & 
   \colhead{$\degr$} & 
   \colhead{ks} 
}

\startdata

% name             HDFv2   RA2000         DEC2000        mi814   PA      Exp(Ks)   
%	       	  	     	              	          
hd4\_0259\_1947 &  4-916 & 12:36:38.605  & 62:12:33.83  & 25.080 & 110, 26.0  & 21.6  \\
hd4\_1076\_1847 &  4-878 & 12:36:40.963  & 62:12:05.30  & 24.009 & 26.0, 137.0  & 24.5  \\
C4-09           &  4-858 & 12:36:41.245  & 62:12:03.07  & 24.904 & 98.0, 136.9   & 21.6  \\
hd4\_0818\_1037 &  4-445 & 12:36:44.640  & 62:12:27.39  & 24.586 & 190.0, 26.0  & 24.5 \\
hd4\_0298\_0744 &  4-316 & 12:36:45.087  & 62:12:50.81  & 24.983 & 145.0  & 24.5 \\
C4-06           &4-555.1 & 12:36:45.409  & 62:11:53.18  & 24.190 & 43.0  & 25.5  \\
hd4\_1006\_0680 &  4-289 & 12:36:46.947  & 62:12:26.08  & 26.082 &   73.0  & 9.9 \\
hd4\_1486\_0880 &  4-382 & 12:36:46.951  & 62:12:05.34  & 24.957 &   13.0, 26.0  & 25.5 \\
hd4\_0367\_0266 &  4-52  & 12:36:47.720  & 62:12:55.79  & 25.579 &   26.0, 108.0  & 21.6 \\
hd2\_1928\_1041 &  2-454 & 12:36:48.266  & 62:14:18.42  & 25.081 &   165.0  & 25.5 \\
C2-05           &  2-449 & 12:36:48.338  & 62:14:16.63  & 24.249 &   165.0  & 25.5  \\
hd2\_1739\_1258 &  2-585 & 12:36:49.811  & 62:14:15.18  & 24.399 &   6.5, 148.0  & 38.9  \\
hd2\_1398\_1164 &  2-525 & 12:36:50.123  & 62:14:01.03  & 25.596 &   83.0  & 9.9 \\
C3-02           &  3-550 & 12:36:51.335  & 62:12:27.51  & 25.706 &   174.0  & 25.5 \\
hd2\_0698\_1297 &  2-604 & 12:36:52.451  & 62:13:37.84  & 25.787 &   42.0, 58.0  & 41.9 \\
hd2\_0705\_1366 &  2-637 & 12:36:52.760  & 62:13:39.09  & 25.916 &   42.0, 58.0  & 41.9 \\ 
hd2\_0529\_1567	&  2-751 & 12:36:54.205  & 62:13:35.83  & 27.233 &   28.9, 31.0  & 38.9 \\
hd2\_0624\_1688 &  2-824 & 12:36:54.627  & 62:13:41.37  & 26.139 &   28.9, 31.0  & 38.9 \\
hd2\_0725\_1818 &  2-903 & 12:36:55.077  & 62:13:47.00  & 25.406 &   28.9, 31.0   & 50.0 \\
hd2\_0743\_1844 &  2-916 & 12:36:55.184  & 62:13:48.07  & 26.214 &   28.9, 31.0   & 50.0 \\
hd3\_1633\_1909 &  3-853 & 12:36:58.967  & 62:12:23.42  & 25.708 &   77.0   & 21.6 \\
hd3\_1824\_1945 &  3-875 & 12:37:00.127  & 62:12:25.22  & 24.671 &   77.0   & 21.6 \\

\enddata
\tablenotetext{a}{Target names are as in \citet{low97} and are in one
of two forms: (1) hd$n$\_xxxx\_yyyy, where $n$ is 2, 3, or 4 and
represents the WFPC2 CCD chip in which the source falls, and xxxx and
yyyy are the pixel coordinates of the source on that chip in the HDF
Version 1 catalog, which was used for original target selection; or, (2)
for some of the brightest targets, C$n$-0$m$ from \citet{ste96b}.
Targets are listed in order of increasing RA.}
\tablenotetext{b} {Source name from HDF Version 2 catalog of
\citet{wil96}, matched to our target by positional proximity and
visual confirmation via the HDF images.}
\tablenotetext{c} {J2000 equatorial coordinates of each target from the HDF
Version 2 catalogs \citet{wil96}.}
\tablenotetext{d} {$I_{814}$ AB isophotal magnitude from the HDF
Version 1 catalog, used for original target selection.}
\tablenotetext{e} {Position angle(s) (degrees east of north) of slitlet
used to observe target.}
\tablenotetext{f} {Total integration time on target at all PAs.}
\end{deluxetable}

%%%%%%%%%%%%%%%%%%%%%%%

\begin{deluxetable}{llrrr}
\tabletypesize{\tiny}
%\rotate
\tablecaption{Unused Kinematic Targets in the {\em Hubble Deep Field.}\label{tab:unusedtargs}}
\tablewidth{0pt}
\tablehead{ 
   \colhead{Name\tablenotemark{a}} & 
   \colhead{HDFv2 \tablenotemark{b}} & 
   \colhead{RA (2000)\tablenotemark{c}} &
   \colhead{DEC (2000)\tablenotemark{c}} & 
   \colhead{$I_{814,AB}$\tablenotemark{d}} \\

   \colhead{~} & 
   \colhead{~} & 
   \colhead{~} &
   \colhead{~} & 
   \colhead{mag} \\
}

\startdata

hd4\_0460\_1146  & 4-491 & 12:36:43.253 & 62:12:38.85 & 25.680  \\
hd4\_1994\_1406  & 4-631 & 12:36:45.310 & 62:11:38.47 & 26.269 \\
hd2\_1949\_0599 &  2-239 & 12:36:45.886  & 62:14:12.09  & 25.342  \\
hd4\_1341\_0299  & 4-85  & 12:36:49.515 & 62:12:20.11 & 26.103  \\
hd3\_0408\_0684 &  3-243 & 12:36:49.814 & 62:12:48.80  & 25.985  \\
hd2\_0664\_0879 & 2-373  &  12:36:50.302 & 62:13:29.73  & 25.951  \\
hd2\_1410\_1282 &  2-594 &  12:36:50.683 & 62:14:03.15  & 26.098 \\
hd3\_0378\_1536 & 4-677 & 12:36:51.620 & 62:12:17.31 & 25.979 \\
hd3\_0457\_2023  & 3-915 &  12:36:53.107 & 62:12:00.74 & 25.770 \\
hd2\_0434\_1377 &  2-643 & 12:36:53.422 & 62:13:29.53  & 25.457 \\
hd3\_1455\_0430 &  3-118 &  12:36:54.727 & 62:13:14.72  & 25.141  \\

\enddata
\tablenotetext{a}{Target names are as in \citet{low97} and are in one
of two forms: (1) hd$n$\_xxxx\_yyyy, where $n$ is 2, 3, or 4 and
represents the WFPC2 CCD chip in which the source falls, and xxxx and
yyyy are the pixel coordinates of the source on that chip in the HDF
Version 1 catalog, which was used for original target selection; or, (2)
for some of the brightest targets, C$n$-0$m$ from \citet{ste96b}.
Targets are listed in order of increasing RA.}
\tablenotetext{b} {Source name from HDF Version 2 catalog of
\citet{wil96}, matched to our target by positional proximity and
visual confirmation via the HDF images.}
\tablenotetext{c} {J2000 equatorial coordinates of each target from the HDF
Version 2 catalogs \citet{wil96}.}
\tablenotetext{d} {$I_{814}$ AB isophotal magnitude from the HDF
Version 1 catalog, used for original target selection.}
\end{deluxetable}

%%%%%%%%%%%%%%%%%%%%%%%%%%%%%%%%%%%%%%%%%%%%%%

\begin{deluxetable}{lcccl}
\tabletypesize{\tiny}
\tablecaption{Redshifts and Spectral Features\label{tab:zs}}
\rotate
\tablewidth{0pt}
\tablehead{ 
   \colhead{Name\tablenotemark{a}} & 
   \colhead{$z$\tablenotemark{b}} & 
   \colhead{$Q_z$\tablenotemark{c}} & 
   \colhead{Feature(s)\tablenotemark{d}}  & 
   \colhead{Remarks} \\
   \colhead{~} & 
   \colhead{~} & 
   \colhead{~} & 
   \colhead{~}  & 
   \colhead{~} \\

}
\startdata
% name             z     Q_z   Feature(s)    comments
%	       			     	              	          
hd4\_0259\_1947 & 0.207 & 2   & -- & new $z$ \\
hd4\_1076\_1847 & $1.010 < z < 2.17$ & 1   & --  & Possible \mgii\ abs \\
C4-09           & 3.226 & 4  &  \lya & quadruple knot \\
hd4\_0818\_1037 & (2.500) & 1   &  -- & -- \\
hd4\_0298\_0744 & $<2.2$ & 1 &  -- & --  \\
C4-06           & 2.794 & 4  & \siii, \oi, \cii, \siiv  & ``Hot Dog''\\
%hd2\_1949\_0599 & (2.427) & 0  &  -- & -- \\
hd4\_1006\_0680 & 2.969 & 3  &  \lya & double knot and em. peak \\
hd4\_1486\_0880 & $<2.29$ & 1 &  -- & --  \\
hd4\_0367\_0266 & 2.931 & 3   &  -- & --  \\
hd2\_1928\_1041 & 2.005 & 4 &  \siii, \oi, \cii, \siiv, \civ, \alii &  C2-06; pair with C2-05\\
C2-05           & 2.008 & 4 &  \siii, \oi, \cii, \siiv, \civ, \alii & pair with hd2\_1928\_1041 \\
hd2\_1739\_1258 & 0.897 & 2   &  -- & new $z$ \\
hd2\_1398\_1164 & 2.236 & 3  &  \lya &  \\
C3-02           & 2.316 & 2  &  -- & $z=2.775$ unlikely; new $z$ \\
hd2\_0698\_1297 & 3.439 & 4   &  \lya & pair with hd2\_0705\_1366  \\
hd2\_0705\_1366 & 3.370 & 4   &  \lya & pair with hd2\_0698\_1297  \\
hd2\_0529\_1567	& 2.268 & 4   &  \lya & new $z$; pair with hd2\_0624\_1688  \\
hd2\_0624\_1688 & 2.424 & 4   &  \lya & pair with hd2\_0529\_1567  \\
hd2\_0725\_1818 & 2.232 & 4   &  \oi, \siii, \alii & pair with hd2\_0743\_1844  \\
hd2\_0743\_1844 & 2.230  & 3   &  \oi, \siii, \alii & new $z$; pair with hd2\_0725\_1818  \\
hd3\_1633\_1909 & 4.056 & 3 & \lya &   \\
hd3\_1824\_1945 & $0.75 < z < 2.2$ & 1 & -- &  \\

\enddata

\tablenotetext{a} {Targets are listed in order of increasing RA, as in
Table~\ref{tab:targs}.}
\tablenotetext{b} {Redshift derived from \lya\ emission line
wavelength or stellar and/or interstellar absorption line wavelengths
(for $z > 1$) or optical nebular emission lines (for $z<1$).  Typical
uncertainties are 0.001 in $z$.  Values in parenthesis indicate
redshifts published by others that we were unable to confirm or
refute.  For details on redshifts listed as ranges or limits, see
text.}
\tablenotetext{c} {Redshift quality.  $0=$ no spectrum obtained;
$1=$highly uncertain, no definitive features identified; $2=$real features are evident but the
redshift is not secure; $3=$redshift probable, with two or more spectral features identified;
$4=$redshift definite, with multiple spectral features identified.}
\tablenotetext{d} {Spectroscopic feature(s) used for measuring
dynamical mass.  \lya\ is in emission only; all others are in
absorption only.}

\end{deluxetable}

%%%%%%%%%%%%%%%%%%%%%%%%%%%%%%%%%%%%%%

\begin{deluxetable}{lcccccccl}
\tabletypesize{\tiny}
\tablecaption{Dynamical Mass Estimates \label{tab:mass}}
\rotate
\tablewidth{0pt}
\tablehead{ 
   \colhead{~}  & 
   \multicolumn{4}{c}{Mass from \lya\ Linewidth} &
   \multicolumn{4}{c}{Mass from Extended Emission} \\
   \colhead{Name\tablenotemark{a}} & 
   \colhead{$v (FWHM)$\tablenotemark{b}}  & 
   \colhead{$r_{\rm 1/2}$\tablenotemark{c}} &
   \colhead{$r_{\rm 1/2}$\tablenotemark{d}} & 
   \colhead{$M_{\rm dyn}$\tablenotemark{e}} &
   \colhead{$\Delta v$\tablenotemark{f}}  & 
   \colhead{$r_{\rm dyn}$\tablenotemark{g}} &
   \colhead{$r_{\rm dyn}$\tablenotemark{h}} & 
   \colhead{$M_{\rm dyn}$\tablenotemark{i}} \\
   \colhead{~}  & 
   \colhead{($\kms$)}  & 
   \colhead{(arcsec)} &
   \colhead{($h^{-1}$ kpc)} & 
   \colhead{($h^{-1} \msun$)} &
   \colhead{($\kms$)}  & 
   \colhead{(arcsec)} &
   \colhead{($h^{-1}$ kpc)} & 
   \colhead{($h^{-1} \msun$)} \\

}
\startdata
% name        v_fwhm rh(``) rh(kpc) M_dyn delta_v  r_dyn(``) r_dyn(kpc)  M_vir  
%	       			     	              	          
hd4\_0259\_1947  & -- & -- & -- & -- & -- & --  & --  & -- \\
hd4\_1076\_1847  & -- & -- & -- & -- & -- & -- & -- & -- \\
C4-09            & $<135$ &  0.14 &  1.0 & $<4.3\exp{9}$  & 820 & $<1.2$ & $<8.5$ & $<1.3\exp{12}$ \\
            & $290$ &  0.14 &  1.0 & $1.9\exp{10}$  & -- & -- & -- & -- \\
hd4\_0818\_1037  & -- & -- & -- & -- & -- & -- & -- & -- \\
hd4\_0298\_0744  & -- & -- & -- & -- & -- & - & -- & -- \\
C4-06            & -- & -- & -- & -- & $<60$ & 3.2 & 24.8 & $<2\exp{10}$ \\
%hd2\_1949\_0599  & -- & -- & -- & -- &  -- & -- & - & -- \\
hd4\_1006\_0680  & 300 &  0.22 &  1.7 & $3.4\exp{10}$  & 677 & 0.3 & 2.3 & $2.4\exp{11}$ \\
hd4\_1486\_0880  & -- & -- & -- & -- & -- & - & -- & -- \\
hd4\_0367\_0266  & -- & -- & -- & -- & -- & - & -- & -- \\
hd2\_1928\_1041  & -- & -- & -- & -- & 390 & 2.0 & 16.5 & $5.6\exp{11}$ \\
C2-05            & -- & -- & -- & -- & 390 & 2.0 & 16.5 & $5.6\exp{11}$ \\
hd2\_1739\_1258  & -- & -- & -- & -- & -- & -- & -- & -- \\
hd2\_1398\_1164  & $<240$ &  0.26 &  2.2 & $<2.8\exp{10}$  & -- & -- & -- & -- \\
C3-02            & -- & -- & -- & -- & -- & -- & -- & -- \\
hd2\_0698\_1297  & $<350$ &  0.45 &  3.3 & $<9.0\exp{10}$  & 4707 & 2.7 & 19.6 & $9.8\exp{13}$ \\
hd2\_0705\_1366  & $<350$ &  0.14 &  1.0 & $<2.8\exp{10}$  & 4707 & 2.7 & 19.8 & $9.8\exp{13}$ \\
hd2\_0529\_1567  & $<270$ &  0.12 &  1.0 & $<1.6\exp{10}$  & 1394 & 6.3 & 50.7 & $2.2\exp{13}$ \\
hd2\_0624\_1688  & $<275$ &  0.31 &  2.5 & $<4.2\exp{10}$  & 1394 & 6.3 & 50.1 & $2.2\exp{13}$ \\
hd2\_0725\_1818  & -- & -- & -- & -- & $<60$ & 1.3 & 10.6 & $<8.6\exp{9}$ \\
hd2\_0743\_1844  & -- & -- & -- & -- & $<60$ & 1.3 & 10.6 & $<8.6\exp{9}$ \\
hd3\_1633\_1909  & 330 &  0.64 &  4.5 & $1.1\exp{11}$  & -- & -- & -- & -- \\
hd3\_1824\_1945  & -- & -- & -- & -- & -- & -- & -- & -- \\

\enddata

\tablenotetext{a} {Targets are listed in order of increasing RA, as in
Table~\ref{tab:targs}.}
\tablenotetext{b} {FWHM of \lya\ emission line measured from the 1D
spectrum and deconvolved with the spectral resolution $FWHM=5$\AA.
Typical uncertainties are $30~\kms$.}
\tablenotetext{c} {Half-light radius as 
measured in the HDF images.  Uncertainties are typically 0\farcs05.}
\tablenotetext{d} {As in $c$, but transformed from arcseconds to kpc
using our adopted flat cosmology with $\Omega_m=0.3,
\Omega_{\Lambda}=0.7, H_0 = 72~ h~ \kmsmpc$.  Uncertainties are $\sim~0.4$ kpc.}
\tablenotetext{e} {Dynamical mass calculated from $M_{\rm dyn} =
r_{\rm 1/2} {v_{FWHM}}^2 / {\rm G}$.  Uncertainties (based solely on
uncertainties in $v_{FWHM}$ and $r_{1/2}$ and neglecting uncertainties
due to non-virial motions, extended stellar or dark matter components,
source inclination, outflows, etc.) on the values in this column are
$\sim 10^{8} \msun$.}
\tablenotetext{f} {Observed velocity spread of
kinematic feature(s).  For close pairs, the velocity separation is
entered twice, once for each component.  Typical uncertainties are
$30~\kms$.}
\tablenotetext{g} {Effective spatial size, spread, or separation as
measured in the HDF images.  Uncertainties are typically 0\farcs1.}
\tablenotetext{h} {As in $g$, but transformed from arcseconds to kpc
using our adopted flat cosmology with $\Omega_m=0.3,
\Omega_{\Lambda}=0.7, H_0 = 72~ h~ \kmsmpc$.  Uncertainties are
$\sim~1$ kpc.}
\tablenotetext{i} {Dynamical mass calculated from $M_{\rm dyn} =
r_{\rm dyn} {\Delta v}^2 / {\rm G}$.  For close pairs, the mass
represents the mass of the system and is entered twice, once for each
component.  Uncertainties (based solely on uncertainties in $\Delta v$
and $r_{\rm dyn}$ and neglecting uncertainties due to non-virial motions,
extended stellar or dark matter components, source inclination,
outflows, etc.) on the values in this column are $\sim 2\exp{8}$.}

\end{deluxetable}

%%%%%%%%%%%%%%%%%%%%%%%%%%%%%%%%%%%%%%%%

\begin{deluxetable}{lcccl}
\tabletypesize{\tiny}
\tablecaption{Stellar Masses from Papovich et al (2005) \label{tab:m*}}
%\rotate
\tablewidth{0pt}
\tablehead{ 
   \colhead{Name\tablenotemark{a}} & 
   \colhead{HNM ID\tablenotemark{b}}  & 
   \colhead{$r$\tablenotemark{c}} &
   \colhead{$z_{\rm used}$\tablenotemark{d}} & 
   \colhead{$M_{\rm *}$\tablenotemark{e}} \\
   \colhead{~}  & 
   \colhead{~}  & 
   \colhead{(arcsec)} &
   \colhead{~} & 
   \colhead{($\msun$)} \\
}
\startdata
% name  HNMID  sep z_used M*
%	       			     	              	          
  hd4\_0259\_1947   &          748 &  0.01 & -1.000 & 0.00 \\
  hd4\_1076\_1847   &         1078 &  0.02 &  0.882 & 2.20E9 \\
          C4-09     &       1114 &  0.57 &  3.216 & 7.84E8 \\
            &       1115 & 0.13 & 3.216 & 3.56E9 \\
  hd4\_0818\_1037   &          989 & 0.04 & 1.444 & 6.68E9 \\
  hd4\_0298\_0744   &          782 & 0.54 & 1.761 & 4.21E9 \\
          C4-06     &       1357 & 0.06 & 2.803 & 1.1E10 \\
            &       1358 & 1.28 & 2.803 & 4.60E9 \\
%  hd2\_1949\_0599   &           97 & 0.02 & 2.427 & 9.49E8 \\
%            &           98 & 0.98 & 2.427 & 1.53E9 \\
  hd4\_1006\_0680   &         1063 & 0.07 & 2.969 & 4.16E8 \\
  hd4\_1486\_0880   &         1282 & 0.02 &-1.000 &  0.00 \\
  hd4\_0367\_0266   &          813 & 0.04 & 2.931 & 7.34E9 \\
            &          814 & 1.06 & 2.931 & 1.8E10 \\
  hd2\_1928\_1041   &          109 & 0.23 & 2.009 & 2.81E9 \\
          C2-05     &        110 & 0.27 & 2.005 & 9.70E9 \\
  hd2\_1739\_1258   &          162 & 0.68 & 1.980 & 1.37E9 \\
            &          163 & 0.41 & 1.980 & 2.3E10 \\
  hd2\_1398\_1164   &          274 & 0.25 & 2.237 & 4.75E9 \\
          C3-02     &       1179 & 0.23 & 2.218 & 1.25E9 \\
%          C3-02     &       1180  0.84 -1.000  0.00
  hd2\_0698\_1297   &          553 & 0.35 & 3.430 & 2.09E9 \\
  hd2\_0705\_1366   &          516 & 0.13 & 3.368 & 1.77E9 \\
  hd2\_0529\_1567   &          616 & 0.08 & 2.161 & 2.50E8 \\
  hd2\_0624\_1688   &          561 & 0.10 & 2.419 & 8.61E8 \\
  hd2\_0725\_1818   &          503 & 0.16 & 2.233 & 1.47E9 \\
 % hd2\_0743\_1844   &          502  0.29  2.617 1.59E9
  hd2\_0743\_1844   &          503 & 1.30 & 2.233 & 1.47E9 \\
  hd3\_1633\_1909   &         1499 & 0.48 & 2.175 & 4.88E8 \\
            &         1506 & 0.99 & 4.050 & 1.36E9 \\
  hd3\_1824\_1945   &         1513 & 0.38 & 2.050 & 4.15E9 \\
\enddata

\tablenotetext{a} {Targets are listed in order of increasing RA, as in
Table~\ref{tab:targs}.}
\tablenotetext{b} {Identification number(s) of object(s) in HDF NICMOS
Mosaic catalog of \citet{pap05} with coordinates matching those of
target galaxy.} 
\tablenotetext{c} {Separation in arcseconds between our target
coordinates and those of matching HNM source.}
\tablenotetext{d} {Redshift used by \citet{pap05} to fit stellar
population model.  A value of -1.000 indicates that no spectroscopic or
photometric redshift was available.}
\tablenotetext{e} {Stellar mass from \citet{pap05}.  A value of 0.00
indicates that no spectroscopic or photometric redshift was
available.}

\end{deluxetable}

\normalsize

\clearpage

%%%%%%%%%%%%%%%%%%%%%%%%%%%%%%%
%
%    FIGURES
%
%%%%%%%%%%%%%%%%%%%%%%%%%%%%%%

\begin{figure}
\epsscale{1.0}
%\plotone{/home/james/hizkin/HDFcomb/rgbmos.v1.ps}
\plotone{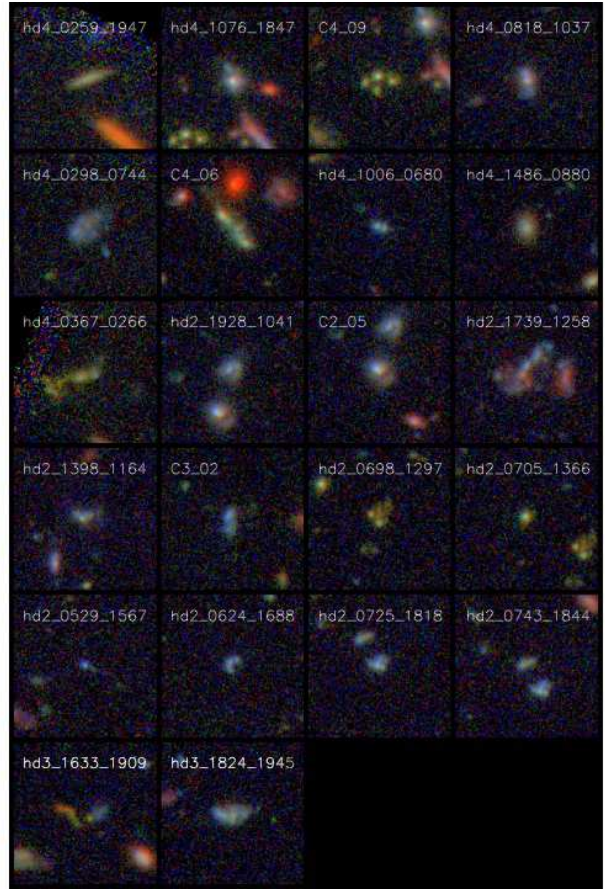}
\caption{Color images of all 22
candidate and confirmed LBGs in our kinematic sample made from
HST/WFPC2 $B_{450}, V_{606}$, and $I_{814}$ HDF images controlling
blue, green, and red pixels, respectively.  Targets are shown in order
of increasing RA, as in Table~\ref{tab:targs}.  North is up, east is
left, and each box is 6\arcsec\ on a side.}
\label{fig:rgb}
\end{figure}

\begin{figure}
\epsscale{0.7}
%\plotone{/home/james/hizkin/HDFcomb/imslitmos_1.v5.eps}
\plotone{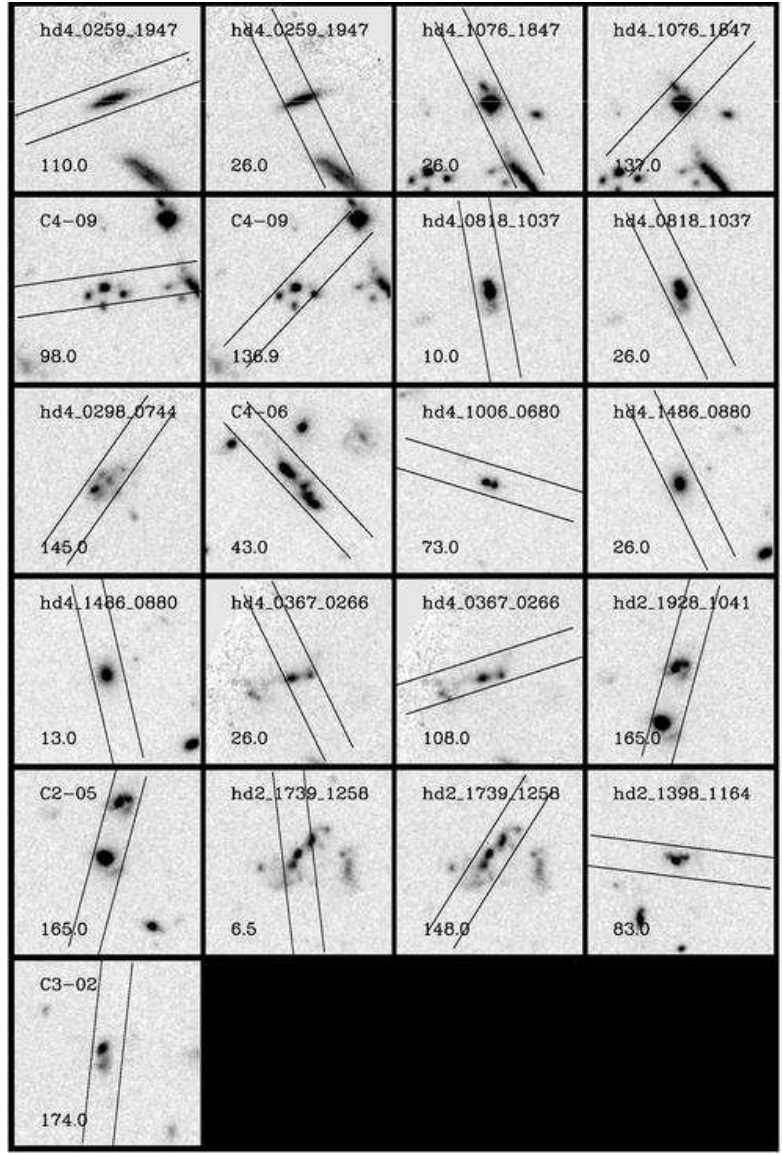}
\caption{HST/WFPC2 $V_{606}$ images of all 22
candidate and confirmed LBGs in our kinematic sample.  A schematic
slitlet 1\farcs1 $\times$ 6\arcsec\ long is shown in each case at the
position angle of the observation, which is labeled in the lower left
of each frame.  Targets are shown in order of increasing RA, as in
Table~\ref{tab:targs}.  North is up, east is left, and each box is
6\arcsec\ on a side.}
\label{fig:images}
\end{figure}
\clearpage
\epsscale{0.7}
{\plotone{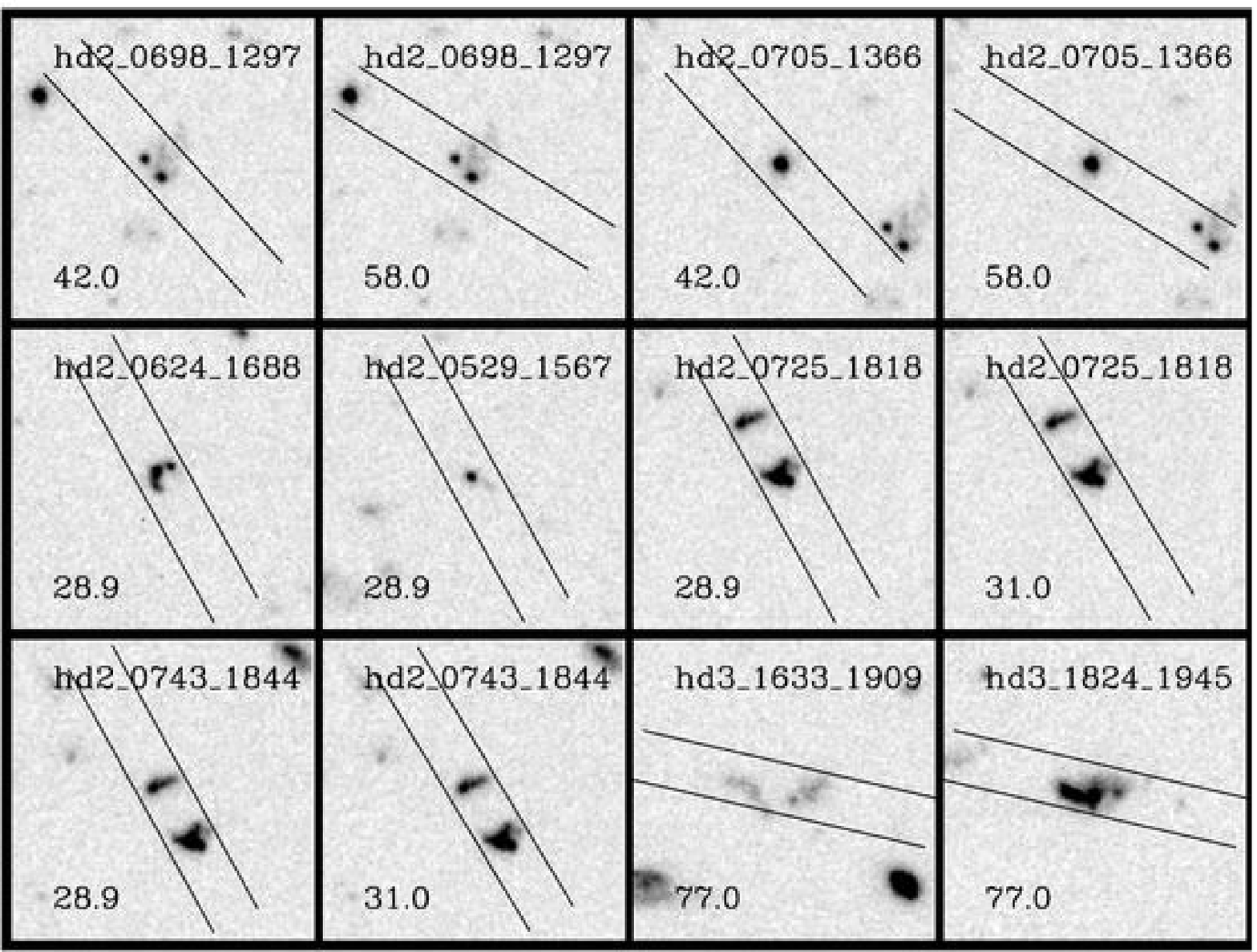}}\\
\centerline{Fig. 2. --- Continued.}
\clearpage

\begin{figure}
\includegraphics[scale=0.5,keepaspectratio=true]{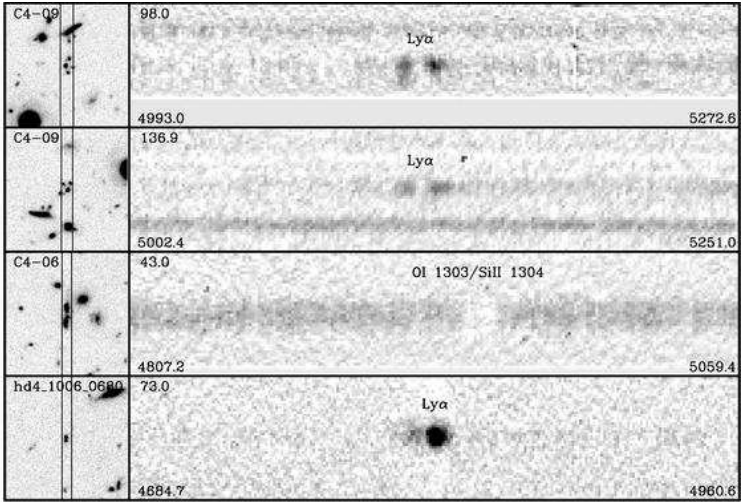}
\caption{HST $I_{814}$ images (left) and closeups of extended kinematic
features in 2D spectra (right) of the 11 LBGs in seven systems
showing such features.  When a pair of targets is covered by a single
slit, both targets are labeled in the image.  Targets are shown in
order of increasing RA, as in Table~\ref{tab:targs}.  Each image is
10\arcsec\ on a side and is oriented to show the slit vertical for
comparison with the 2D spectrum.  The slit width is 1\farcs1~ $\times
{\rm cos} i$, where $i$ is the relative position angle of the slitlet
with respect to the slit mask.  Each spectrum covers the same 10\arcsec\
vertically as the image, and roughly 250\AA\ horizontally, the exact
value depending on the projected pixel scale and therefore, again, the
position angle of the particular slitlet used relative to its parent
slit mask.  The beginning and ending wavelengths (\AA, lower left and
right of spectrum) of each section of spectrum and the position angle
of the slitlet ($\degr$ east of north, upper left of spectrum) are
labeled.  The features include spatially resolved or separated pairs
of absorption lines of \alii,
\siii, and \oi; and spatially or spectrally resolved or separated
pairs of emission lines of \lya.  The emission lines are usually the
only relevant kinematic feature in their parent spectra; the
absorption lines however are chosen as representative from among
several.}
\label{fig:2dA}
\end{figure}
\clearpage
\centerline{\includegraphics[scale=0.5,keepaspectratio=true]{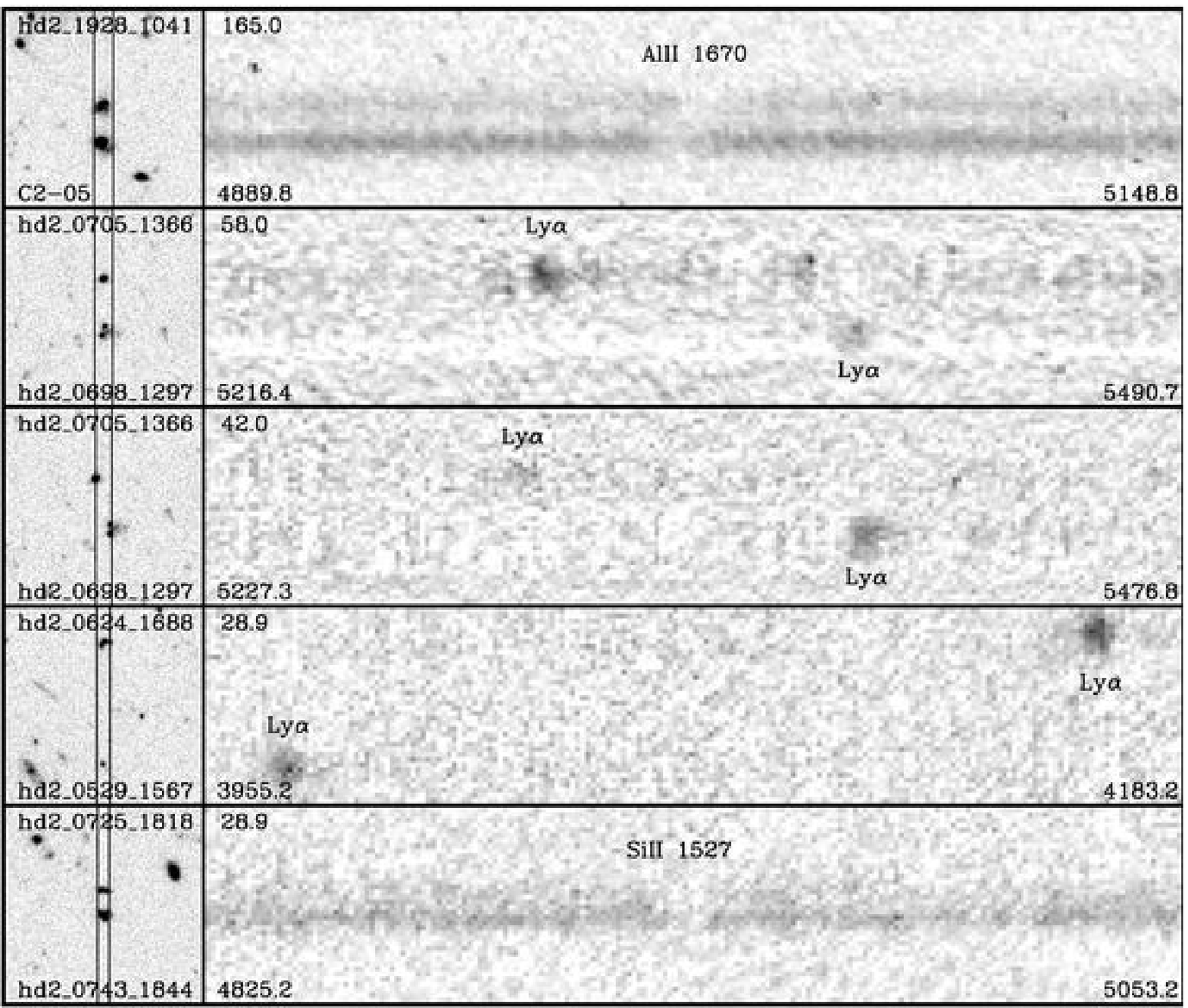}}
\ \\[5mm]
\centerline{Fig. 3. --- Continued.}
\clearpage

\begin{figure}
%\plotone{/home/james/hizkin/HDFcomb/plot1d_a.v2.eps}
\plotone{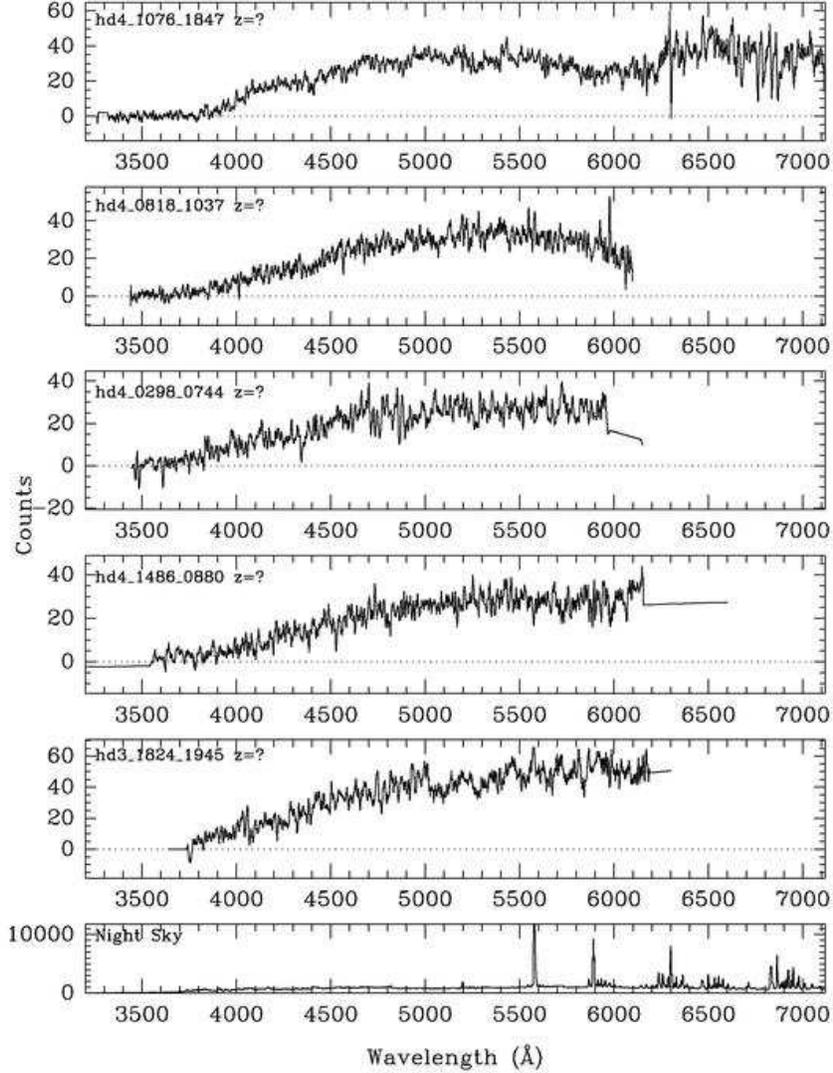}
\caption{One-dimensional spectra extracted from our two-dimensional
LRIS spectra of all 22 candidate and confirmed LBG kinematic targets.
Spectra are shown in order of increasing redshift, except in the first
panel, for which redshifts are unknown (redshift quality $Q_z=1$) and
the objects are shown in order of increasing RA.  Redshifts with
$Q_z=2$ are followed by '?'.  Each panel includes
a spectrum of the night sky showing strong teluric emission lines.
The wavelengths of common LBG emission and absorption lines are marked
on each spectrum.  The spectra have been cleaned of strong night sky
emission line residuals and smoothed by 7 pixels ($\sim 8$\AA) to
suppress noise.  }
\label{fig:1da}
\end{figure}
\clearpage
{\plotone{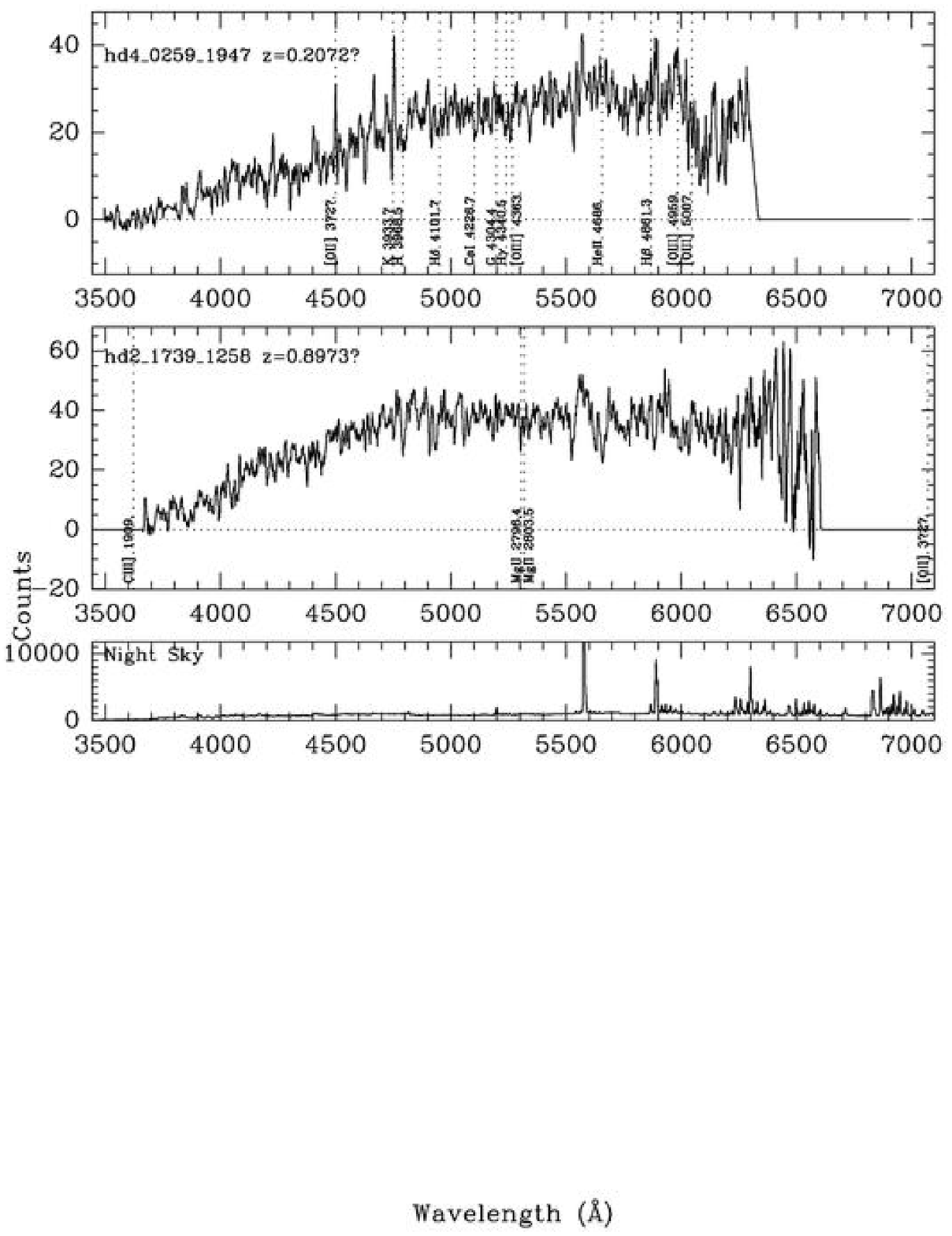}}\\
\centerline{Fig. 4. --- Continued.}
\clearpage
{\plotone{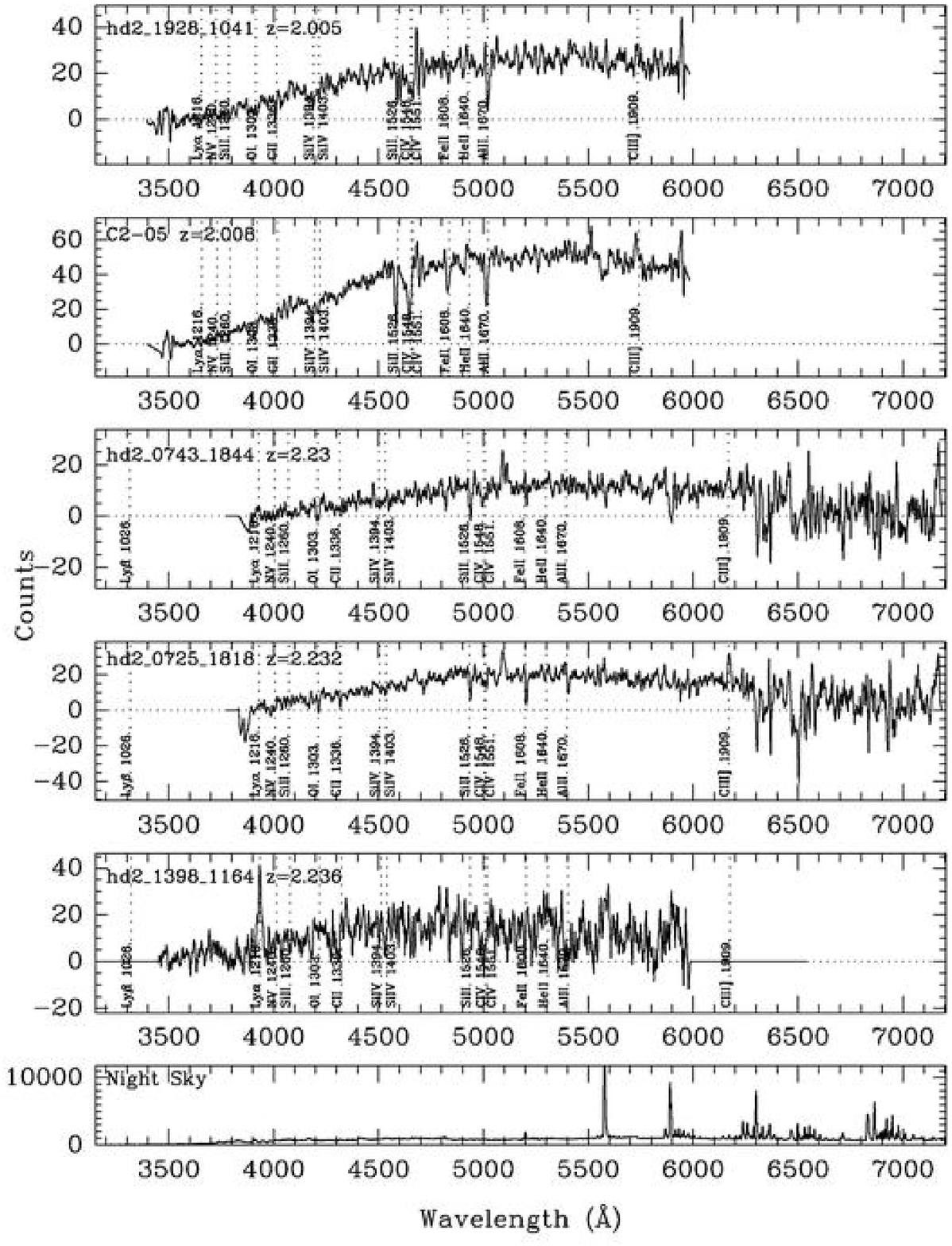}}\\
\centerline{Fig. 4. --- Continued.}
\clearpage
{\plotone{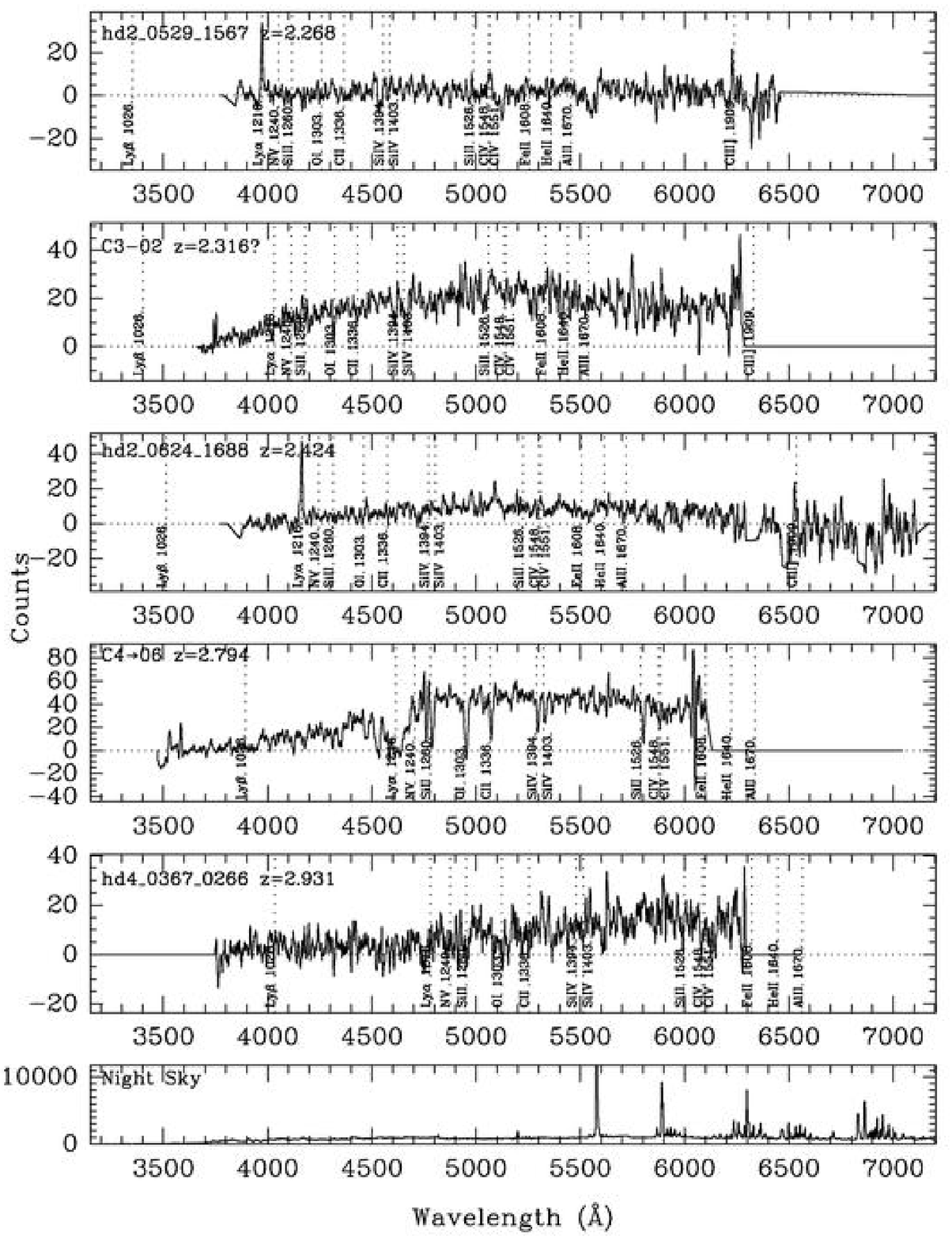}}\\
\centerline{Fig. 4. --- Continued.}
\clearpage
{\plotone{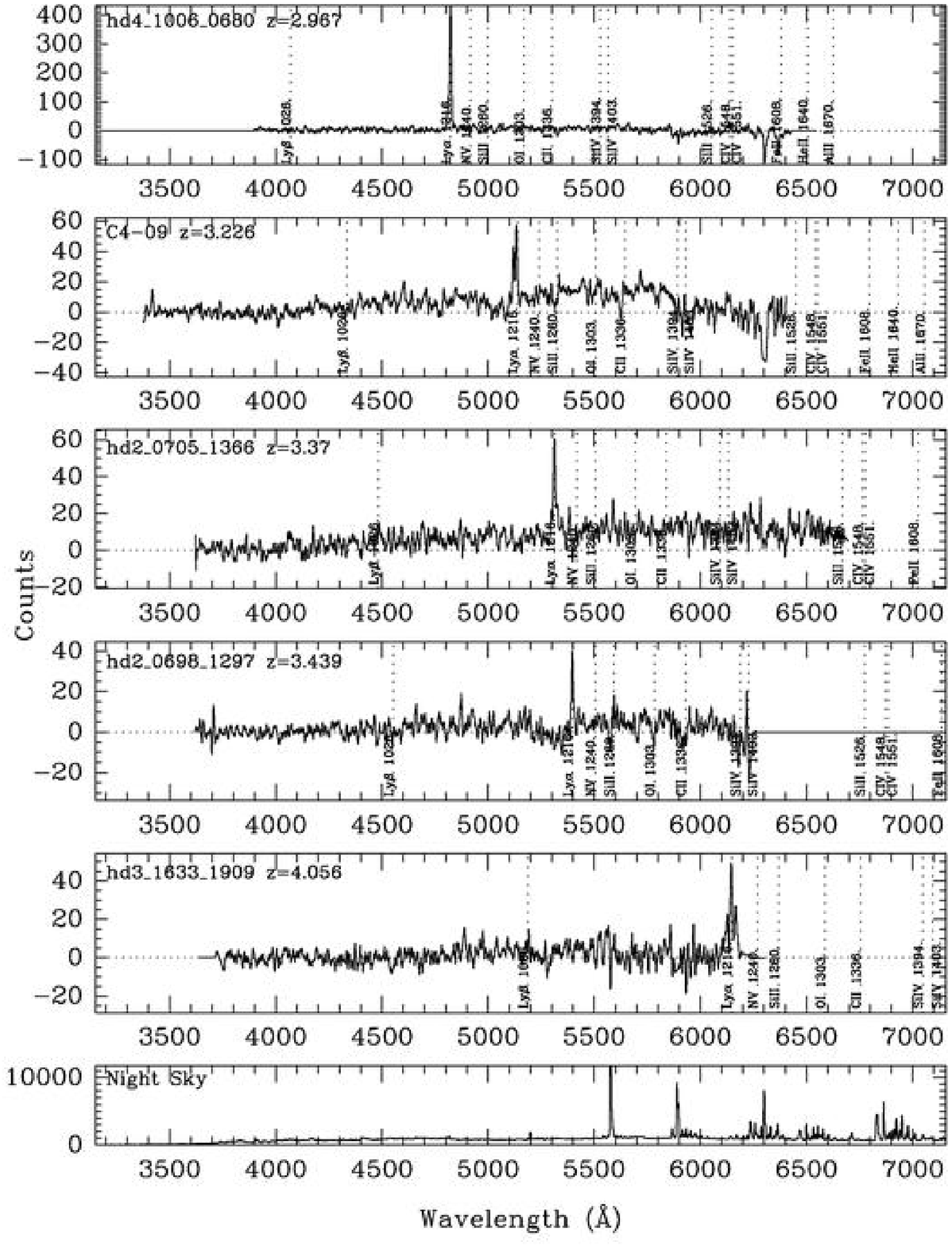}}\\
\centerline{Fig. 4. --- Continued.}
\clearpage

\begin{figure}
\epsscale{1.0}
%\plotone{sep_dv.eps}
\plotone{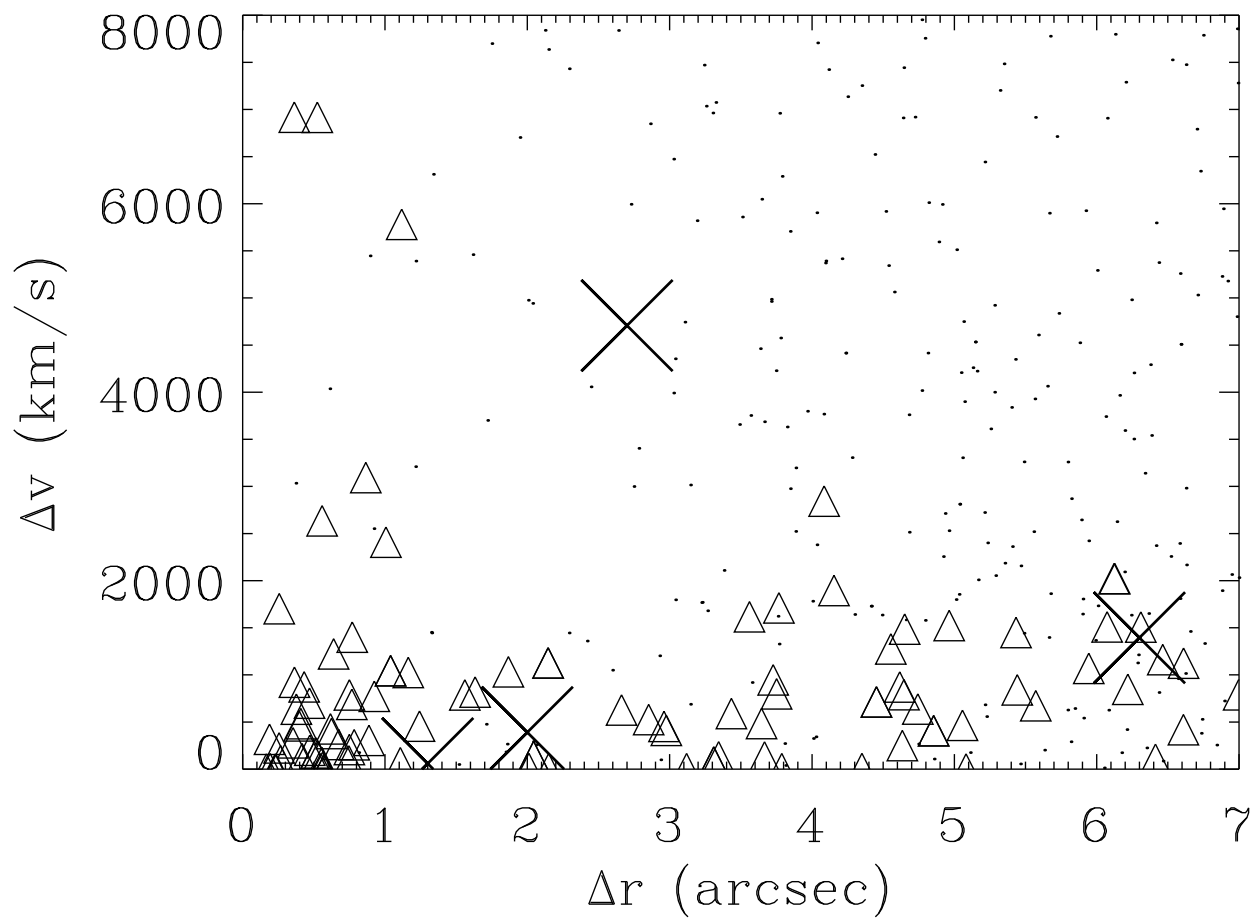}
\caption{Pairwise velocities vs. projected separation for close pairs of simulated LBGs with projected separation $\Delta r<7$\arcsec\ for LBGs in Model 3 of \citet{kamp07}.  Open triangles indicate pairs with cosmological redshift difference $\Delta z < 0.001$, while dots indicate close pairs with larger redshift differences.  The values for the four close pairs in our sample are shown with a $\times$ symbol.}
\label{fig:deltav}
\end{figure}

\begin{figure}
\epsscale{1.0}
%\plotone{mh_mv.eps}
\plotone{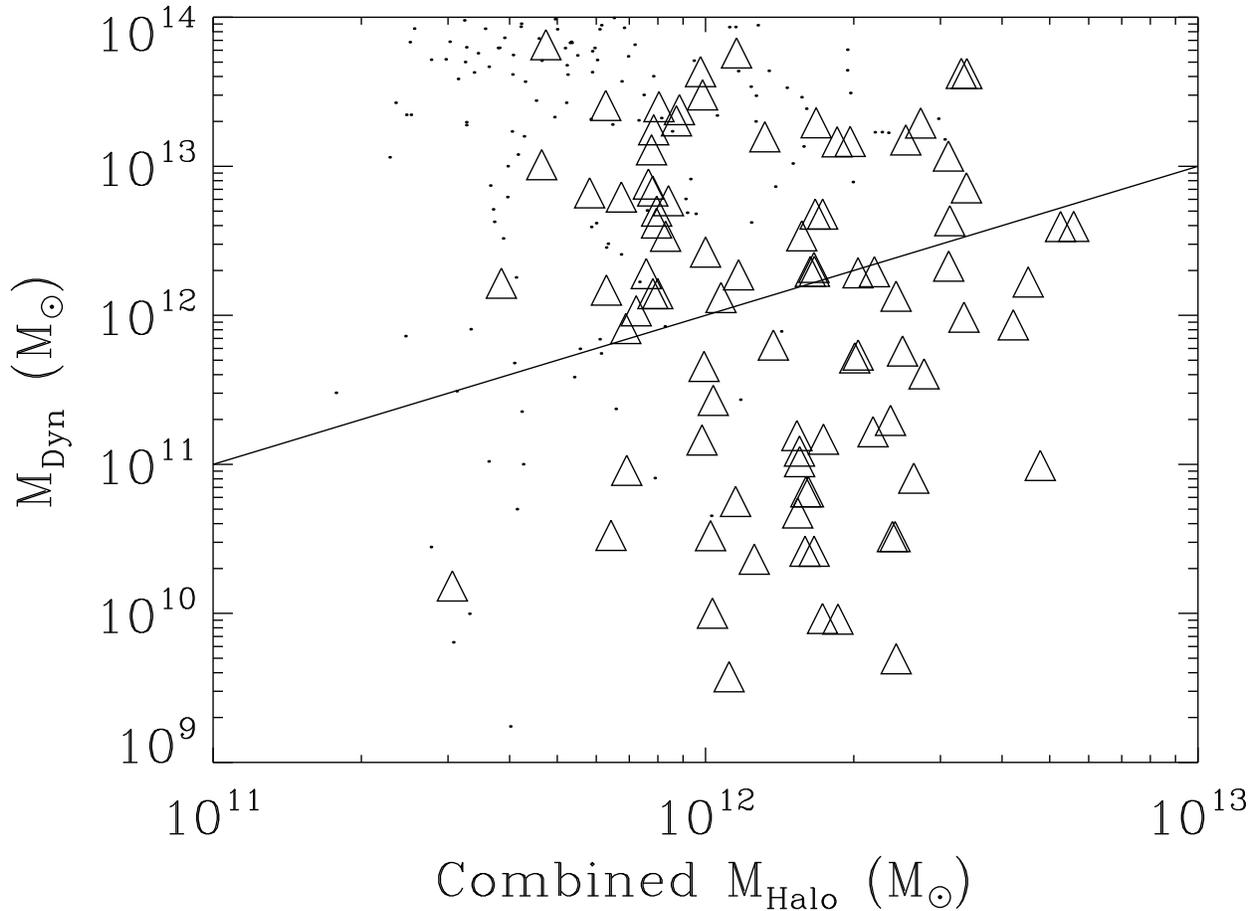}
\caption{Calculated dynamical mass (from Eq.~\ref{eq:mdyn}) vs. combined halo mass for close pairs of LBGs with projected separation $\Delta r<7$\arcsec\ in the simulation of \citet{kamp07}.  Symbols are as in Fig.~\ref{fig:deltav}.   The solid line represents $M_{\rm Halo} = M_{\rm Dyn}$.  No strong correlation is seen.}
\label{fig:mhmv}
\end{figure}

\begin{figure}
\includegraphics[angle=-90,scale=0.5,keepaspectratio=true]{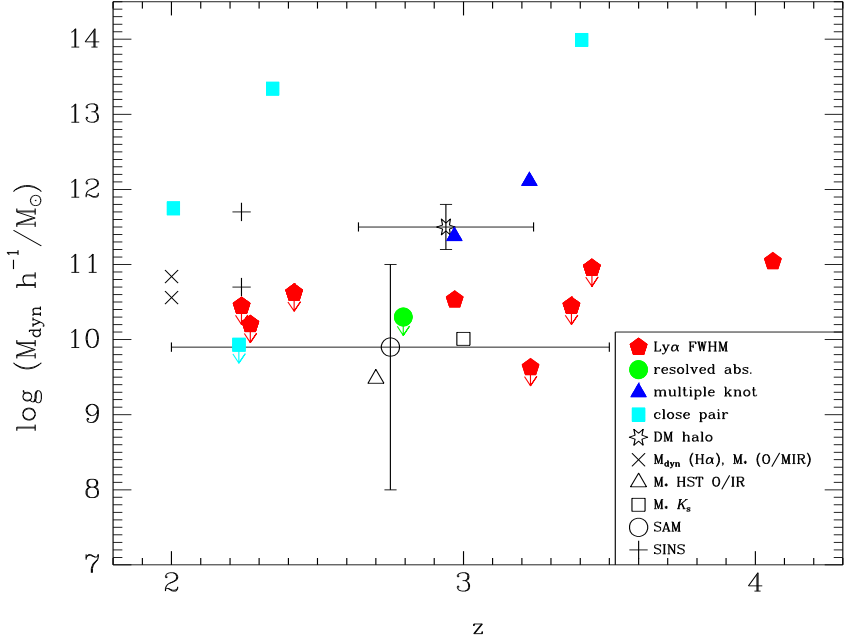}
\caption{Estimated dynamical mass vs. redshift for our 13 LBGs with kinematic
signatures.  Masses of close pair systems are shown as light blue
filled squares at the pair's average redshift, multiple knots as dark
blue filled triangles, the galaxy (C4-06) with resolved absorption as
a green filled circle, and masses derived from \lya\ linewidths as red
filled pentagons.  Also plotted are the mean value for the stellar masses
derived by \citet{pap01} using HST/WFPC2 + NICMOS optical and NIR
photometry ({\em open triangle}); the median value for stellar masses
of luminous LBGs by \citet{sha01} using ground-based NIR ($J$ and
$K_s$) photometry ({\em open square}); the mean value for the dark
matter halos of LBGs at $z=2.9$ derived using clustering analysis by
\citet{ade05a} ({\em star symbol}); the mean dynamical mass of
BM/BX galaxies at $z\sim2$ measured using redshifted
\ha\ emission line widths and the stellar mass using optical-MIR
photometry including from {\it Spitzer} by \citet{erb06} (upper and
lower {\em crosses}, respectively); the mean dynamical mass of BM/BX
galaxies in the SINS survey measured with VLT/SINFONI by \citet{for06}
({\it plus symbol}); and the theoretical median and range predicted by
the semi-analytical models of \citet{som01} (\em open circle).  }
\label{fig:Mz}
\end{figure}

\begin{figure}
\includegraphics[angle=-90,scale=0.3,keepaspectratio=true]{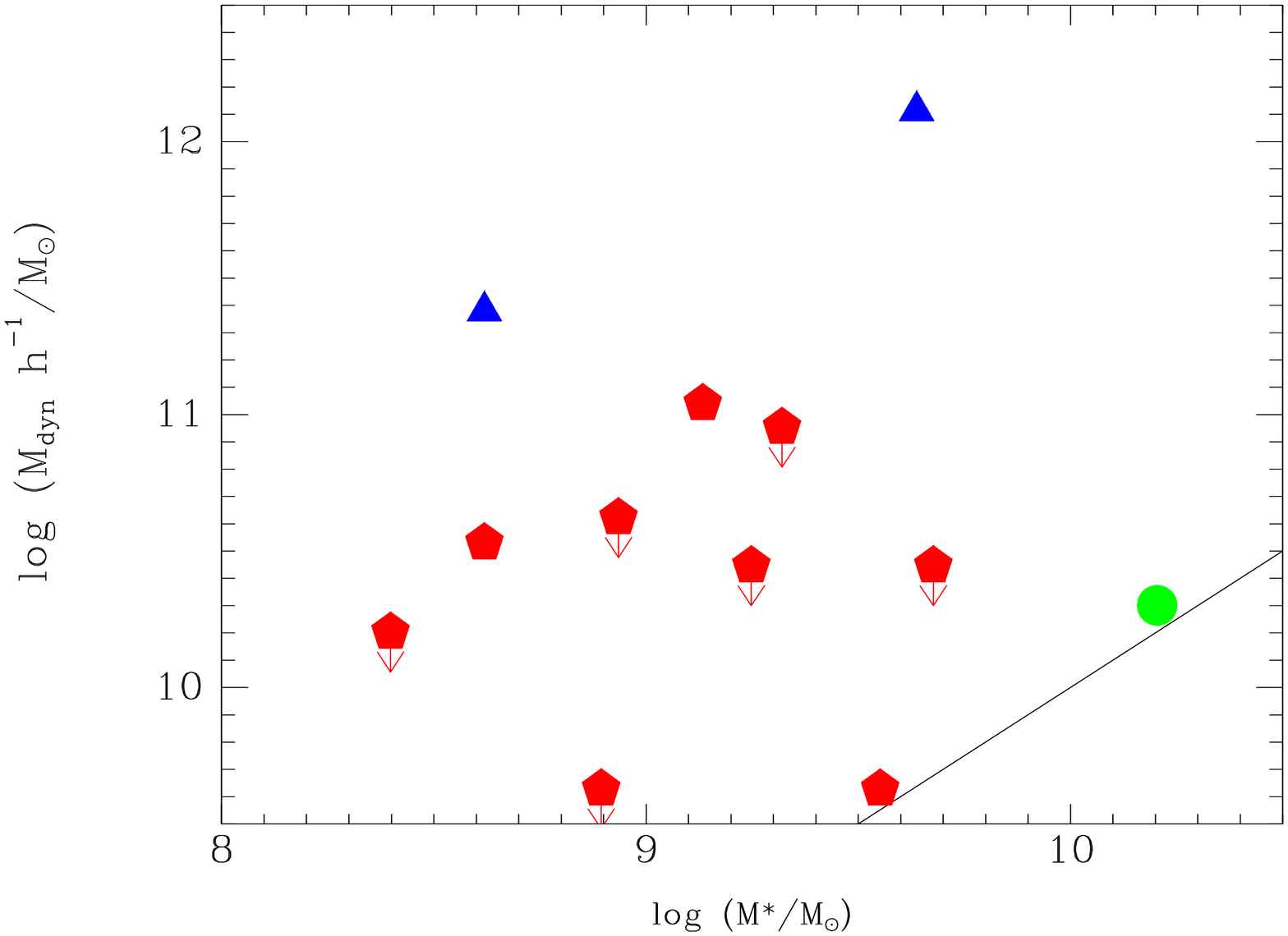}
\includegraphics[angle=-90,scale=0.3,keepaspectratio=true]{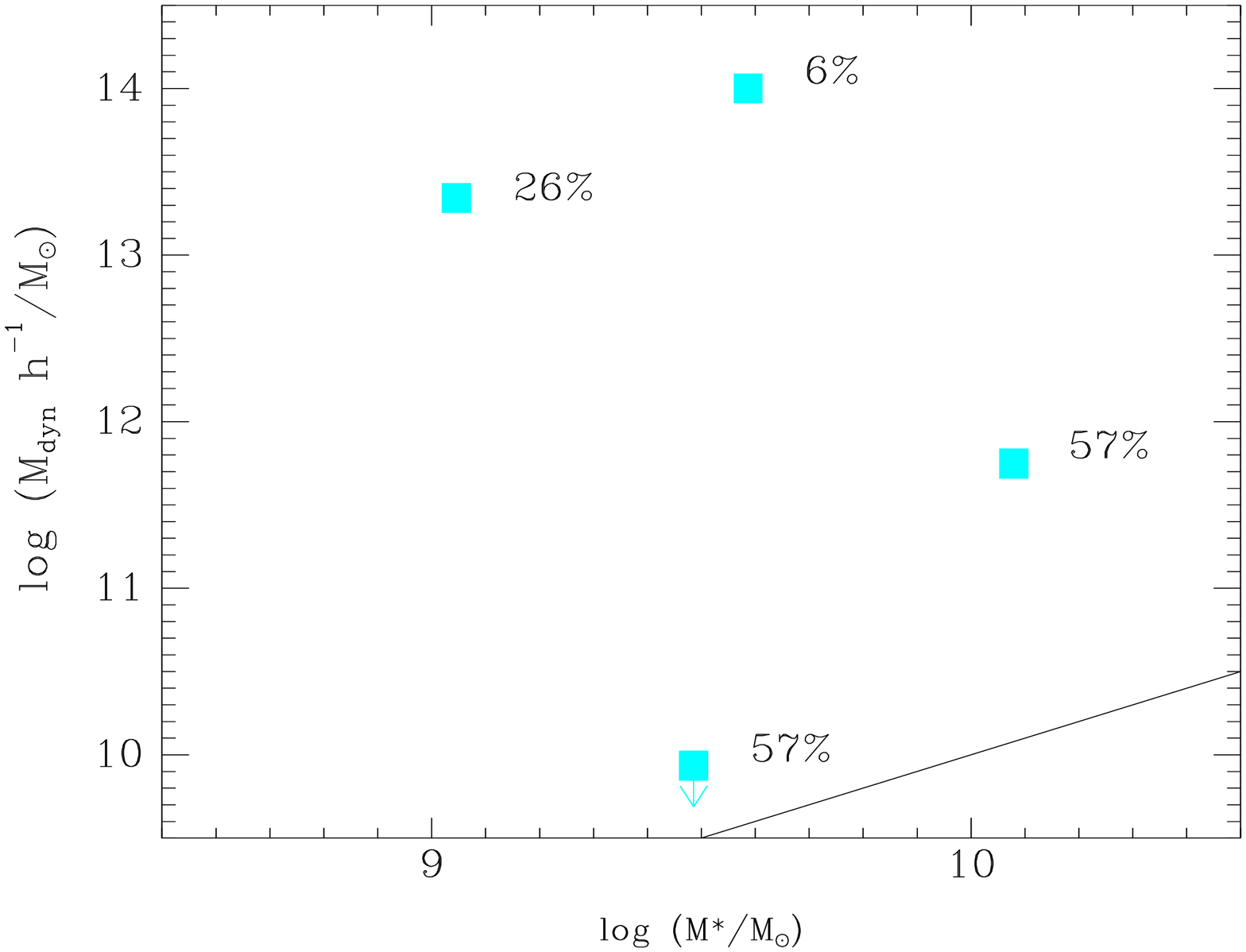}
\caption{({\it left}) Stellar mass vs. estimated dynamical mass for individual
LBGs with kinematic signatures.  ({\it right}) Combined stellar mass
vs. estimated system dynamical mass for close-pair systems.  Stellar masses are
from \citet{pap05}.  Symbols are as in Fig.~\ref{fig:Mz}, and the
diagonal lines represent $M* = M_{\rm dyn}$.  Each point is labeled with the probability that the corresponding pair is a true physical association rather than a chance superposition, as estimated from the simulation results shown in Fig.~\ref{fig:deltav}.}
\label{fig:m*}
\end{figure}

\clearpage

\end{document}